\documentclass[aps,pra,amsmath,showpacs,amssymb]{revtex4}

\usepackage{graphicx}
\usepackage{color}

\newcommand{\be}{\begin{equation}}
\newcommand{\ee}{\end{equation}}

\newcommand{\bex}{\begin{eqnarray}}
\newcommand{\eex}{\end{eqnarray}}

\newtheorem{thm}{Theorem}
\begin{document}

\title{Entanglement, intractability and no-signaling}
\author{R. Srikanth}
\email{srik@ppisr.res.in}
\affiliation{Poornaprajna Institute of Scientific Research,
Sadashivnagar, Bangalore- 560080, India.}
\affiliation{Raman Research Institute, Sadashivnagar,
Bangalore- 560080, India.}

\begin{abstract}
We consider  the problem of  deriving the no-signaling  condition from
the assumption that, as  seen from a complexity theoretic perspective,
the universe is not an exponential place. A fact that disallows such a
derivation is  the existence  of {\em polynomial  superluminal} gates,
hypothetical primitive  operations that enable  superluminal signaling
but not the efficient  solution of intractable problems.  It therefore
follows, if  this assumption is  a basic principle of  physics, either
that it  must be supplemented with additional  assumptions to prohibit
such  gates, or,  improbably,  that no-signaling  is  not a  universal
condition.  Yet, a gate of  this kind is possibly implicit, though not
recognized  as  such,  in  a  decade-old  quantum  optical  experiment
involving  position-momentum entangled  photons.  Here  we  describe a
feasible  modified  version  of the experiment  that  appears  to  explicitly
demonstrate the  action of this gate. Some  obvious counter-claims are
shown to  be invalid.  We  believe that the unexpected  possibility of
polynomial  superluminal operations  arises  because some  practically
measured quantum  optical quantities  are not describable  as standard
quantum mechanical observables.
\end{abstract}
\maketitle \date{}

\section{Introduction}
In  a multipartite quantum  system, any  completely positive  (CP) map
applied  locally to  one  part  does not  affect  the reduced  density
operator of the remaining part.  This fundamental no-go result, called
the  ``no-signalling  theorem''   implies  that  quantum  entanglement
\cite{epr}  does  not  enable  nonlocal  (``superluminal'')  signaling
\cite{nos}  under standard  operations,  and is  thus consistent  with
relativity,  inspite of the  counterintuitive, stronger-than-classical
correlations  \cite{bell}  that   entanglement  enables.   For  simple
systems,  no-signaling follows  from  non-contextuality, the  property
that the probability assigned to  projector $\Pi_x$, given by the Born
rule, Tr$(\rho\Pi_x)$, where $\rho$  is the density operator, does not
depend   on    how   the   orthonormal   basis    set   is   completed
\cite{gle57,ash95}.   No-signaling has  also been  treated as  a basic
postulate to derive quantum theory \cite{gis01halv}.

It is of interest to  consider the question of whether/how computation
theory,  in particular intractability  and uncomputability,  matter to
the  foundations of (quantum)  physics. Such  a study,  if successful,
could  potentially  allow  us  to   reduce  the  laws  of  physics  to
mathematical  theorems about  algorithms and  thus shed  new  light on
certain  conceptual  issues.   For   example,  it  could  explain  why
stronger-than-quantum   correlations    that   are   compatible   with
no-signaling  \cite{bra06} are disallowed  in quantum  mechanics.  One
strand of thought  leading to the present work,  earlier considered by
us  in Ref.  \cite{sri06},  was the  proposition that  the measurement
problem is  a consequence of basic algorithmic  limitations imposed on
the computational power that can be supported by physical laws. In the
present work,  we would like to  see whether no-signaling  can also be
explained  in  a  similar  way, starting  from  computation  theoretic
assumptions.

The central problem in
computer science  is the conjecture that  two computational complexity
classes, {\bf  P} and  {\bf NP}, are  distinct in the  standard Turing
model  of computation.  ${\bf P}$  is the  class of  decision problems
solvable in  polynomial time by  a (deterministic) TM.  ${\bf  NP}$ is
the class  of decision problems  whose solution(s) can be  verified in
polynomial time  by a deterministic TM.   $\#{\bf P}$ is  the class of
counting problems associated with the decision problems in ${\bf NP}$.
The word  ``complete" following a  class denotes a problem  $X$ within
the class, which is maximally hard in the sense that any other problem
in the  class can be  solved in poly-time  using an oracle  giving the
solutions of  $X$ in a  single clock cycle.  For  example, determining
whether  a Boolean  forumla  is satisfied  is  {\bf NP}-complete,  and
counting   the   number    of   Boolean   satisfactions   is   $\#{\bf
P}$-complete. The word ``hard" following a class denotes a problem not
necessarily  in the  class, but  to which  all problems  in  the class
reduce in poly-time.

{\bf P} is often taken to be the class of computational problems which
are  ``efficiently solvable"  (i.e., solvable  in polynomial  time) or
``tractable", although  there are potentially larger  classes that are
considered tractable such as {\bf  RP} \cite{srirp} and {\bf BQP}, the
latter being the class of  decision problems efficiently solvable by a
quantum  computer \cite{srirp}.   {\bf  NP}-complete and  potentially
harder problems, which  are not known to be  efficiently solvable, are
considered intractable in the Turing model.  If ${\bf P} \ne {\bf NP}$
and the universe is a polynomial-- rather than an exponential-- place,
physical  laws cannot  be harnessed  to efficiently  solve intractable
problems, and  {\bf NP}-complete problems  will be intractable  in the
physical world.


That classical physics supports  various implementations of the Turing
machine is  well known. More  generally, we expect  that computational
models supported by a physical  theory will be limited by that theory.
Witten identified  expectation values  in a topological  quantum field
theory with values  of the Jones polynomial that  are $\#{\bf P}$-hard
\cite{wit89}.   There  is  evidence  that  a physical  system  with  a
non-Abelian topological  term in  its Lagrangian may  have observables
that are {\bf NP}-hard, or even \#{\bf P}-hard \cite{mic98}.

Other recent  related works that have studied  the computational power
of  variants  of  standard  physical  theories from  a  complexity  or
computability      perspective      are,      respectively,      Refs.
\cite{cal04,insel,aar05,bram98,sri06}  and  Refs.  \cite{cal04,sri06}.
Ref.  \cite{aar05}  noted that {\bf NP}-complete problems  do not seem
to  be  tractable  using  resources  of  the  physical  universe,  and
suggested that  this might embody a  fundamental principle, christened
the  {\bf  NP}-hardness  assumption  (also cf.   \cite{sciam}).   Ref.
\cite{srigru}  studies how  insights from  quantum  information theory
could be used to constrain physical laws.  We will informally refer to
the  proposition  that the  universe  is  a  polynomial place  in  the
computational  sense  (to  be  strengthened  below)  as  well  as  the
communication sense by the expression  ``the world is not hard enough"
(WNHE)  \cite{sriwnhe}. 

Recently, Ref.  \cite{ben09} have posed the question
 whether nonlinear quantum
evolution can be considered as providing any help in solving otherwise
hard  problems, on  the grounds  that under  nonlinear  evolution, the
output  of such a  computer on  a mixture  of inputs  is not  a convex
combination of  its output on the  pure components of  the mixture. We
circumvent  this  problem here  by  adopting  the  standpoint of  {\em
  information  realism},   the  position  that   physical  states  are
ultimately  information states registered  in some  way sub-physically
but objectively by Nature. At this  stage, we will not worry about the
details except to note an implication for the present situation, which
is that  from the  perspective of `Nature's  eye', there are  no mixed
states. Therefore, in describing nonlinear physical laws or specifying
the working of non-standard computers  based on such laws, it suffices
for our purpose  to specify their action on  (all relevant) pure state
inputs.

In Ref.  \cite{sri06}, we pointed out that the assumption of WNHE (and
further that of {\bf P} $\ne$ {\bf NP}) can potentially give a unified
explanation of (a) the observed `insularity-in-theoryspace' of quantum
mechanics (QM),  namely that QM  is {\em exactly} unitary and  
linear, and
requires  measurements   to  conform  to  the   $|\psi|^2$  Born  rule
\cite{sriw,insel}; (b) the classicality  of the macroscopic world; (c)
the lack of quantum physical mechanisms for non-signaling superquantum
correlations \cite{bra06}.

In (a),  the basic idea  is that departure  from one or more  of these
standard  features  of  QM  seems  to invest  quantum  computers  with
super-Turing power to solve hard problems efficiently, thus making the
universe   an  exponential   place,  contrary   to   assumption.   The
possibility (b) arises for the  following reason.  It is proposed that
the WNHE assumption holds not only in the sense that hard problems (in
the  standard  Turing  model)  are  not efficiently  solvable  in  the
physical  world,   but  in  the  stronger  sense   that  any  physical
computation  can be simulated  on a  probabilistic TM  with at  most a
polynomial slowdown  in the number of steps  (the Strong Church-Turing
thesis). Therefore,  the evolution of  any quantum system  computing a
decision problem, could asymptotically be simulated in polynomial time
in the size of  the problem, and thus lies in {\bf  BPP}, the class of
problems  that  can  be  efficiently  solved  by  a  probabilistic  TM
\cite{sribpp}.

Assuming {\bf  BPP} $\ne$  {\bf BQP}, this  suggests that  although at
small  scales,  standard QM  remains  valid  with characteristic  {\bf
  BQP}-like behavior,  at sufficiently large  scales, classical (`{\bf
  BPP}-like') behavior should emerge, and that therefore there must be
a  definite scale-- sometimes  called the  Heisenberg cut--  where the
superposition   principle   breaks    down   \cite{bag00},   so   that
asymptotically, quantum states are  not exponentially long vectors. In
Ref.   \cite{sri06}, we  speculate that  this  scale is  related to  a
discretization of  Hilbert space.   This approach provides  a possible
computation theoretic  resolution to the  quantum measurement problem.
In (c),  the idea  is that  in a polynomial  universe, we  expect that
phenomena in which  a polynomial amount of physical  bits can simulate
exponentially   large   (classical)   correlations,   thereby   making
communication complexity trivial, would be forbidden.

In  the  present work,  we  are  interested  in studying  whether  the
no-signaling theorem follows from the WNHE assumption.  The article is
divided into two parts: Part I, concerned with the computer scientific
aspects, giving  a complexity theoretic motivation for  the work; Part
II,  concerned  with the  quantum  optical  implementation  of a  test
suggested by Part I.

In Part I, first  some results concerning non-standard operations that
violate no-signaling  and help efficiently  solve intractable problems
are  surveyed,  in  Sections  \ref{sec:srisum}  and  \ref{sec:sripol},
respectively.  Then, in Section \ref{sec:sripolynon}, we introduce the
concept of  a polynomial  superluminal gate, a  hypothetical primitive
operation that  is prohibited by  the assumption of  no-signaling, but
allowed if instead we only assume that intractable problems should not
be  efficiently  solvable  by  physical  computers.   We  examine  the
relation between the above two classes of non-standard gates.  We also
describe a {\em  constant} gate on a single  qubit or qutrit, possibly
the simplest instance of a polynomial superluminal operation.

In  Part II, first  we present  a quantum  optical realization  of the
constant  gate,  and  its   application  to  an  experiment  involving
entangled light generated by  parametric downconversion in a nonlinear
crystal in  Section \ref{sec:ent}. Physicists who could  not care less
about  computational complexity  aspects could  skip directly  to this
Section.  They may  be warned that the intervening  sections of Part I
will  involve mangling QM  in ways  that may  seem awkward,  and whose
consistency  is,  unfortunately,  not  obvious!  On  the  other  hand,
computer scientists  unfamiliar with  quantum optics may  skip Section
\ref{sec:ent},    which   is    essentially    covered   in    Section
\ref{sec:sriung}, which  discusses quantitative and  conceptual issues
surrounding the  physical realization of the  constant gate.  Finally,
we   conclude  with   Section  \ref{sec:conclu}   by   surveying  some
implications  of   a  possible   positive  outcome  of   the  proposed
experiment, and discussing how  such an unexpected physical effect may
fit in with the mathematical structure of known physics.  We present a
slightly   abridged   version  of   discussions   in   this  work   in
Ref. \cite{sriv1}.

\section{Part I: Computation theoretic motivation}

\subsection{Superluminal gates \label{sec:srisum}}

Even minor variants of QM are known to lead to superluminal signaling.
An   example  is   a  variant   incorporating   nonlinear  observables
\cite{ganeshpol91},   unless   the   nonlinearity   is   confined   to
sufficiently  small  scales \cite{srisvet,srisvet0,srisvet1,srisvet2}.
In this Section,  we will review the case  of violation of no-signling
due  to  departure  from  standard  QM via  the  introduction  of  (a)
non-complete  Schr\"odinger evolution  or  measurement, (b)  nonlinear
evolution  \cite{srisvet3},  (c) departure  from  the Born  $|\psi|^2$
rule.

In each  case, we  will not  attempt to develop  a non-standard  QM in
detail,  but   instead  content  ourselves   with  considering  simple
representative examples.
 
(a)  {\em  Non-complete  measurements  or  non-complete  Schr\"odinger
  evolution.}    Let  us  consider   a  QM   variant  that   allows  a
non-tracepreserving    (and   hence   non-unitary)    but   invertible
single-qubit operation of the form:
\begin{equation}
\label{eq:g}
G = \left( \begin{array}{ll} 1 & 0 \\ 0 & 1+\epsilon
\end{array}\right),
\end{equation}
where $\epsilon  > 0$ is a  real number.  The  resultant state $\sum_x
\alpha_x  |x\rangle$  must  be   normalized  by  dividing  it  by  the
normalization factor $\sqrt{\sum_x |\alpha_x|^2}$ immediately before a
measurement, making measurements  nonlinear. Given the entangled state
$(1/\sqrt{2})(|01\rangle +  |10\rangle)$ that Alice and  Bob share, to
transmit a  superluminal signal, Alice applies either  $G^m$ (where $m
\geq 1$  is an integer)  or the identity  operation $I$ to  her qubit.
Bob's  particle  is left,  respectively,  in  the state  $\rho_B^{(1)}
\propto                \frac{1}{2}(|0\rangle\langle0|                +
(1+\epsilon)^{2m}|1\rangle\langle1|)$      or      $\rho_B^{(0)}     =
\frac{1}{2}(|0\rangle\langle0|  + |1\rangle\langle1|)$,  which  can in
principle  be distinguished,  the  distance between  the states  being
greater for  larger $m$ (cf.  Section \ref{sec:sripol}),  leading to a
superluminal signal from Alice to Bob.

More generally,  we may  allow non-unitary and  irreversible evolution
but still  conform to no-signaling, provided the  corresponding set of
operator(s) is {\em complete}, i.e., constitutes a partition of unity.
Suppose Alice and  Bob share the state $\rho_{AB}$,  and Alice evolves
her part of $\rho_{AB}$ locally  through the linear operation given by
the set  ${\cal P}$ of  (Kraus) operator elements $\{E_{j}  \equiv e_j
\otimes\mathbb{I}_B,      j=1,2,3,\cdots\}$     \cite{nc00},     where
$\mathbb{I}_B$  is the  identity  operator in  Bob's subspace.   Bob's
reduced   density  operator   $\rho^{\prime}_B$  conditioned   on  her
performing the operation and after normalization is:
\begin{equation}
\label{nosig}
\rho^{\prime}_B          =         {\cal          N}^{-1}         {\rm
  Tr_A}\left[\sum_{j}E_{j}\rho_{AB}E^{\dag}_{j}\right] = {\cal N}^{-1}
    {\rm          Tr}_A\left[\sum_{j}E^{\dag}_{j}E_{j}\rho_{AB}\right],
    ~~~~{\cal                 N}                 =                {\rm
      Tr}_{AB}\left[\sum_{j}E^{\dag}_{j}E_{j}\rho_{AB}\right],
\end{equation} 
where  ${\cal   N}$  is  the  normalization  factor.    We satisfy the
no-signaling condition
$\rho^\prime_B = \rho_B$ only  if $\rho_{AB}$ is unentangled or ${\cal
  P}$ satisfies the completeness relation
\begin{equation}
\sum_j   e_j^\dag  e_j   = \mathbb{I}_A,
\label{eq:srimadhwa}
\end{equation}
which guarantees  that the operation preserves norm  ${\cal N}$.  Here
$\mathbb{I}_A$ is  the identity operator in Alice's  subspace.  If the
norm  is  not  preserved,  renormalization  is  required,  making  the
evolution  effectively nonlinear. If  the system  $A$ is  subjected to
unitary  evolution  or  non-unitary  evolution  due to  noise,  or  to
standard   projective  measurements   or  more   general  measurements
described by positive operator  valued measures, the corresponding map
satisfies Eq.   (\ref{eq:srimadhwa}), and $\rho_B^{\prime}  = \rho_B$.
For terminological brevity, we call a (non-standard) gate like $G$, or
a  non-complete   operation  ${\cal  P}$   that  enables  superluminal
signaling,  as  `superluminal  gate',   and  denote  the  set  of  all
superluminal  gates  by `$C^{<}$'.   For  the  purpose  of this  work,
$C^{<}$  is   restricted  to  qubit  or   qutrit  gates.   Non-unitary
super-quantum  cloning or  deleting, introduced  in  Ref.  \cite{sen},
which  lead   to  superluminal   signaling,  are  other   examples  of
non-complete operations.

Even at the single-particle level, if the measurement is non-complete,
there   is   a   superluminal    signaling   due   to   breakdown   in
non-contextuality  coming  from  the  renormalization.   As  a  simple
illustration, suppose we are given two observers Alice and Bob sharing
a       delocalized       qubit,      $\cos(\theta/2)|0\rangle       +
\sin(\theta/2)|1\rangle$,  with eigenstate $|1\rangle$  localized near
Alice and $|0\rangle$  near Bob.  With an $m$-fold  application of $G$
(which  can be  thought of  as an  application of  imaginary  phase on
Alice's side, leading to  selective augmentation of amplitude) on this
state,      Alice      produces      the     (unnormalized)      state
\mbox{$\cos(\theta/2)|0\rangle      +       (1      +      \epsilon)^m
  \sin(\theta/2)|1\rangle$},  so  that  after  renormalization,  Bob's
probability    of   obtaining   $|0\rangle$    has   changed    in   a
context-dependent fashion from $\cos^2(\theta/2)$ to $\cos^2(\theta/2)
(\cos^2(\theta/2)  +  (1+\epsilon)^{2m}  \sin^2(\theta/2))^{-1}$.   By
thus  nonlocally  controlling the  probability  with  which Bob  finds
$|0\rangle$, Alice  can probabilistically signal Bob superluminally.

(b)  {\em  Nonlinear  evolution.}   As  a  simple  illustration  of  a
superluminal gate  arising from nonlinear evolution,   we consider the
action of  the nonlinear  two-qubit `OR' gate  $R$, whose action  in a
preferred (say, computational) basis is given by:
\begin{equation}
\label{eq:nonlinor}
\left.
\begin{array}{ll}
|00\rangle \pm |11\rangle \\
|01\rangle \pm |10\rangle \\
|01\rangle \pm |11\rangle \\
\end{array} \right\} \stackrel{R}{\longrightarrow}
|01\rangle \pm |11\rangle; \hspace{1.0cm}
\begin{array}{l}
|00\rangle \pm |10\rangle 
\stackrel{R}{\longrightarrow} |00\rangle \pm |10\rangle), \\
|\alpha\beta\rangle \stackrel{R}{\longrightarrow} |\alpha\beta\rangle.
\end{array}
\end{equation}
If the  two qubits are entangled  with other qubits, then  the gate is
assumed to act in each subspace labelled by states of the other qubits
in the computational  basis.  Alice and Bob share  the entangled state
$|\Psi\rangle = 2^{-1/2}(|00\rangle -  |11\rangle)$. To transmit a bit
superluminally Alice measures her  qubit in the computational basis or
the  diagonal  basis   $\{|\pm\rangle  \equiv  2^{-1/2}(|0\rangle  \pm
|1\rangle\}$,   leaving   Bob's   qubit's   density  operator   in   a
computational basis  ensemble or a diagonal basis  ensemble, which are
equivalent in standard QM.  However, with the nonlinear operation $R$,
the  two ensembles  can be  distinguished. Bob  prepares  an ancillary
qubit in  the state $|0\rangle$,  and applies a  CNOT on it,  with his
system qubit  as the control. On  the resulting state  he performs the
nonlinear  gate  $R$, and  measures  the  ancilla.  The  computational
(resp., diagonal)  basis ensemble yields the value  1 with probability
$\frac{1}{2}$ (resp.,  1). By  a repetition of  the procedure  a fixed
number $m$ of  times, a superluminal signal is  transmitted from Alice
to Bob with exponentially small  uncertainty in $m$.  Analogous to Eq.
(\ref{eq:nonlinor}), one  can define a `nonlinear  AND', which, again,
similarly leads to a nonlocal signaling.

(c)  {\em  Departure  from  the  Born  $|\psi|^2$  probability  rule.}
Gleason's theorem shows that the Born probability rule that identifies
$|\psi|^2$  as a probability  measure, and  more generally,  the trace
rule, is the  only probability prescription consistent in  3 or larger
dimensions     with    the     requirement     of    non-contextuality
\cite{gle57}. Suppose we retain  unitary evolution, which preserve the
2-norm, but assume that the  probability of a measurement on the state
$\sum_j  \alpha_j|j\rangle$   is  of  the   form  $|\alpha_j|^p/\sum_k
|\alpha_k|^p$ for  outcome $j$, and $p$ any  non-negative real number.
The renormalization will make  the measurement contextual, giving rise
to a  superluminal signal.  One might consider  more general evolution
that preserves a  $p$-norm, but there are no  linear operators that do
so except permutation matrices \cite{insel}.

For example,  let Alice  and Bob share  the two-qubit  entangled state
$\cos\theta|00\rangle + \sin\theta|11\rangle$  ($0 < \theta < \pi/2)$.
The probability for Alice  measuring her particle in the computational
basis and finding $|0\rangle$ (resp., $|1\rangle$) must be the same as
that  for a  joint measurement  in  this basis  to yield  $|00\rangle$
(resp.,  $|11\rangle$). Therefore  Bob's reduced  density  operator is
given by  the state  $\rho^{(1)} = (\cos^p\theta  |0\rangle\langle0| +
\sin^p\theta  |1\rangle\langle1|)/(\cos^p\theta +  \sin^p\theta)$.  On
the other hand, if Alice employs an ancillary, third qubit prepared in
the state $|0\rangle$, and applies a Hadamard on it conditioned on her
qubit  being  in the  state  $|0\rangle$,  she  produces the  state  $
\frac{\cos\theta}{\sqrt{2}}|000\rangle                                +
\frac{\cos\theta}{\sqrt{2}}|001\rangle  + \sin\theta|110\rangle$.  The
probability that Alice obtains outcomes 00,  01 or 10 must be that for
a joint measurement to yield 000,  001 or 110.  Along similar lines as
in the above case we find that she leaves Bob's qubit in the state
\begin{equation}
\rho^{(2)} \equiv
\frac{2^{(1-p/2)}\cos^p\theta     |0\rangle\langle0|     +    \sin^p\theta
|1\rangle\langle1|)}
{2^{(1-p/2)}\cos^p\theta  + \sin^p\theta}.  
\end{equation}
Since    $\rho^{(1)}$   and    $\rho^{(2)}$    are   probabilistically
distinguishable, with sufficiently many shared copies Alice can signal
Bob one bit superluminally, unless $p=2$.

\subsection{Exponential gates \label{sec:sripol}}

As  superluminal  quantum  gates   like  $G$  or  $R$  are  internally
consistent, one can  consider why no such operation  occurs in Nature,
whether a  fundamental principle prevents  their physical realization.
One     candidate    principle     is    of     course    no-signaling
itself. Alternatively, since we would  like to derive it, linearity of
QM  may  be taken  as  an axiom.   Since  all  the above  non-standard
operations involve  an overall nonlinear evolution,  the assumption of
strict  quantum   mechanical  linearity  can  indeed   rule  out  such
non-standard  gates. Yet  it  must  be admitted  that,  from a  purely
physics  viewpoint, assuming  that  QM is  linear  affords no  greater
insight than assuming it to  be a non-signaling theory.  We would like
to  suggest  that  the the  absence  of  such  operations may  have  a
complexity theoretic basis.

Both  superluminal gates  as  well as  hypothetical  gates that  allow
efficient  solving  of  intractable  problems  involve  some  sort  of
communication  across  superposition  branches.   In  particular,  the
superluminal gates of Section  \ref{sec:srisum} can be turned into the
latter type of gates, as discussed below.

(a) {\em Non-complete quantum gates.}  It is easily seen that the gate
$G$  in  Eq. (\ref{eq:g})  can  be  used  to solve  {\bf  NP}-complete
problems efficiently.  Consider  solving boolean satisfiability (SAT),
which is {\bf NP}-complete:  given an efficiently computable black box
function $f: \{0,1\}^n \mapsto  \{0,1\}$, to determine if there exists
$x$ such  that $f(x) =  1$.  With the  use of an oracle  that computes
$f(x)$, we prepare the $(n+1)$-qubit entangled state
\begin{equation}
|\Psi_{nc}\rangle = 2^{-n/2}\sum_{x\in\{0,1\}^n}|x\rangle|f(x)\rangle, 
\label{eq:sat}
\end{equation}
and then apply  $G^m$ to the second, 1-qubit register,  where $m$ is a
sufficiently  large  integer,   before  measuring  the  register.   In
particular,   suppose  that   at  most   one  solution   exists.   The
un-normalized  `probability  mass'  of obtaining  outcome  $|1\rangle$
becomes  1  (and the  normalized  probability  about  1/2) when  $m  =
n/(2\log(1+\epsilon))$, if there is a solution, or, if no solution
exists,  remains 0.   Repeating the  experiment a  fixed  number of
times, and applying the Chernoff bound,  we find that to solve SAT, we
only require $m  \in O(n)$.  For terminological brevity,  we will call
as  `exponential  gate' such  a  non-standard  gate  that enables  the
efficient computation of {\bf NP}-complete problems, and denote by $E$
the set  of all exponential gates,  restricted in the  present work to
qubits  and qutrit  gates. 

(b) {\em Nonlinear quantum gates.}  The nonlinear operation $R$ in Eq.
(\ref{eq:nonlinor})    can   be    used   to    efficiently   simulate
nondeterminism.     We   prepare    the   state    $|\psi\rangle$   in
Eq. (\ref{eq:sat}), where the first  $n$ qubits are called the `index'
qubits and the  last one the `flag' qubit.   There are $2^{n-1}$ 4-dim
subspaces, consisting  of the  first index qubit  and the  flag qubit,
labelled by the index qubits $2,\cdots,n$.  On each such subspace, the
first index qubit and flag qubit  are in one of the states $|00\rangle
+ |11\rangle$,  $|01\rangle + |10\rangle$,  $|00\rangle + |10\rangle$.
The operation  Eq.  (\ref{eq:nonlinor}) is applied  $n$ times, pairing
each index qubit sequentially with  the flag. The number of terms with
1 on  the flag bit doubles with  each operation so that  after the $n$
operations, it becomes disentangled and can then be read off to obtain
the  answer \cite{bram98}.   A slight  modification of  this algorithm
solves  $\#{\bf P}$-complete  problems efficiently,  by  replacing the
flag qubit with  $\log n$ qubits and the  1-bit nonlinear OR operation
with the corresponding nonlinear  counting.  The final readout is then
the  number  of solutions  to  $f(x)=1$  \cite{bram98}.  Applying  the
nonlinear OR and AND alternatively  to the state $|\psi\rangle$ in Eq.
(\ref{eq:sat}) allows one to  efficiently solve the quantified Boolean
formula problem, which is {\bf PSPACE}-complete \cite{pspace}.

(c)  {\em  Non-Gleasonian  gates.}   By  employing  polynomially  many
ancillas in the method of (c) in the previous subsection, one can show
that non-Gleasonian quantum computers (for  which $p \ne 2$) can solve
{\bf PP}-complete  problems \cite{sripp} efficiently.   Defining ${\bf
BQP}_p$  as similar  to ${\bf  BQP}$, except  that the  probability of
measuring  a  basis   state  $|x\rangle$  equals  $|\alpha_x|^p/\sum_y
|\alpha_y|^p$ (so  that ${\bf  BQP}_2 = {\bf  BQP}$), it can  be shown
that ${\bf PP} \subseteq {\bf BQP}_p$ for all constants $p \ne 2$, and
that, in particular,  ${\bf PP}$ exactly characterizes the  power of a
quantum computer with even-valued $p$ (except $p=2$) \cite{insel}.

In view  of the connection  between the two  classes of gates,  we now
propose, as we earlier did  in Ref.  \cite{sri06}, that the reason for
the  absence   in  Nature  of   the  superluminal  gates   of  Section
\ref{sec:srisum} is  WNHE: in a  universe that is a  polynomial place,
exponential gates like $G$ and $R$ are ruled out.  In the next Section
we  will  consider  in  further  detail  the  viability  of  the  WNHE
assumption as an explanation for no-signaling.

\subsection{Polynomial superluminal gates \label{sec:sripolynon}}

Even though  WNHE excludes the  type of superluminal  gates considered
above, for the exclusion to be general, it would have to be shown that
every superluminal  gate is exponential, i.e., $C^<  \subseteq E$.  It
turns  out  that  this  cannot  be done,  because  one  can  construct
hypothetical   {\it   polynomial   superluminal  gates},   which   are
superluminal  operations that  are  not exponential.  In  fact, it  is
probably  true that $E  \subset C^<$.   To see  this, let  us consider
solving   the   {\bf   NP}-complete   problem  associated   with   Eq.
(\ref{eq:sat}) via Grover search \cite{srigro97}, which is optimal for
QM \cite{sriben97} but offers  only a quadratic speed-up, thus leaving
the  complexity of the  problem exponential  in $n$,  at least  in the
black box setting.  The optimality proof relies on showing that, given
the problem of distinguishing an empty oracle ($\forall_x A(x)=0$) and
a non-empty  oracle containing a  single random unknown string  $y$ of
known length $n$ (i.e.  $A(y) =  1$, but $\forall_{x\ne y} A(x) = 0$),
subject to the  constraint that its overall evolution  be unitary, and
linear  (so  that  in  a  computation  with  a  nonempty  oracle,  all
computation  paths querying  empty  locations evolve  exactly as  they
would for an empty oracle), the speed-up over a classical search is at
best quadratic.

Any degree  of amplitude amplification  of the marked state  above the
quadratic level would then  require empty superposition branches being
`made  aware'  of  the  presence   of  a  non-empty  branch,  i.e.,  a
nonlinearity  of  some  sort.   Let  us  suppose  Bob  can  perform  a
trace-preserving   nonlinear    transformation   $\rho_j   \rightarrow
\tilde{\rho}_j$ of the above kind  on an unknown ensemble of separable
states.  Further, let Alice and Bob share an entangled state, by which
Alice is able to prepare, employing two different 
positive operator-valued measures (POVMs), two different
but equivalent  ensembles of Bob.  Then, depending  on Alice's choice,
his  reduced density  matrix  evolves as  $\rho_B  = \sum_j  p_j\rho_j
\rightarrow \sum  p_j \tilde{\rho}_j \equiv \rho^\prime$  or $\rho_B =
\sum_s   p_k\rho_k   \rightarrow   \sum  p_k   \tilde{\rho}_k   \equiv
\rho^{\prime\prime}$  where   $(\rho_j,p_j)$  and  $(\rho_k,p_k)$  are
distinct,  equivalent ensembles  \cite{sriper02}.   The assumption  of
linearity    is   sufficient   to    ensure   that    $\rho^\prime   =
\rho^{\prime\prime}$.   This  is not  guaranteed  in  the presence  of
nonlinearity,  leading  to  a  potential superluminal  signal.   In  a
nonlinearity of the above kind, the result would depend on whether the
particular ensemble remotely prepared by Alice has states that include
$|y\rangle$  in the  superposition.   This would  lead  to a  scenario
similar  to  that  encountered  with  nonlinear gate  $R$  in  Section
\ref{sec:srisum}.


Possibly the  simplest examples  of polynomial superluminal  gates are
the non-invertible  {\em constant  gates}, which map  any state  in an
input Hilbert space to a fixed  state in the output Hilbert space, and
have the form $|\xi\rangle \otimes  \sum_j \langle j|$, for some fixed
$\xi$. Examples in matrix notation are:
\begin{equation}
Q =  \left(\begin{array}{ll}   1  &  1  \\  0  &  0
\end{array}\right);\hspace{0.5cm}
Q ^{\prime} 
= \left(\begin{array}{lll} 1 & 1 & 1 \\ 0 & 0 & 0 \\ 0 & 0 & 0
\end{array}\right),
\label{eq:Kjanichwara}
\end{equation}
acting      in     Hilbert      space     ${\cal      H}_2     \equiv$
span$\{|0\rangle,|1\rangle\}$     and      ${\cal     H}_3     \equiv$
span$\{|0\rangle,|1\rangle,|2\rangle\}$, respectively.   They have the
effect of  mapping any input state  in ${\cal H}_2$ to  a fixed (apart
from  a  normalization  factor)  state  $|\xi\rangle$,  in  this  case
$|\xi\rangle$ being $|0\rangle$.  In Eq. (\ref{eq:Kjanichwara}), we do
not in general 
require the input and output bases to be the same, nor indeed that
the input  and output Hilbert subspaces  be the same  (for example, as
with  the  distinct  incoming  and  outgoing  modes  of  a  scattering
problem.)

Both $Q$  and $ Q^{\prime}$  are non-complete, inasmuch  as $Q^{\dag}Q
\ne  \mathbb{I}$ and  $(Q^\prime)^\dag Q^\prime  \ne  \mathbb{I}$, and
represent  superluminal  gates.   For  example,  by  applying  or  not
applying $Q$  to her register in the  state $(1/\sqrt{2})(|01\rangle +
|10\rangle)$ shared with Bob, Alice  can remotely prepare his state to
be the pure state $(1/\sqrt{2})(|0\rangle + |1\rangle)$ or leave it as
a maximal  mixture, respectively, corresponding to a superluminal
signal of about 0.3 bits (determined by the Holevo bound).
Similarly, by choosing  to apply, or
not, $Q^\prime$  on her half  of the state  $(1/\sqrt{2})(|11\rangle +
|22\rangle)$ shared  with Bob, Alice  can superluminally signal  him.

The constant gate is linear and presumes no re-normalization following
its  non-complete  action.  The  probability  of  the  occurance of  a
constant  gate $C$ when  it is  applied to  a state  $|\psi\rangle$ is
simply given by  $||C|\psi\rangle||^2$, per the standard prescription.
One consequence is  that it could not be  used to violate no-signaling
without the use of entanglement.  As an illustration: in ${\cal H}_3$,
let the states $|0\rangle$ and $|1\rangle$ be localized near Alice and
$|2\rangle$   near   Bob.    Applying   $Q^{\prime}$  on   the   state
$|\psi\rangle  \equiv  a|0\rangle +  b|1\rangle  + c|2\rangle$,  Alice
obtains  the  (unnormalized)  state  $Q^{\prime}|\psi\rangle  =  (a  +
b)|0\rangle  + c|2\rangle$.   If renormalization  were  allowed, Alice
could contextually  (i.e., nonlocally) influence  Bob's probability to
find $|2\rangle$  to be $|c|^2/(|a+b|^2+|c|^2)$  or $|c|^2$, depending
on  whether she  applies  $Q^\prime$  or not.   The  linearity of  the
constant   gate  requires  the   interpretation  that   following  her
application  of  $Q^\prime$,  Alice   can  detect  the  particle  with
probability $|a+b|^2$, while for Bob, the probability remains $|c|^2$.
Though   non-complete   operations   do   not   necessarily   conserve
probability, still, as we will find below and later that in situations
of interest they can exactly or effectively conserve probability.

On the other  hand, neither $Q$ nor $Q^\prime$  nor a general constant
gate is an exponential gate:  each of them simply transforms any valid
input into a  fixed output.  Intuitively, this lack  of any dependence
on the  input clearly limits its computational  power.  Operations $Q$
and $Q^\prime$ in Eq. (\ref{eq:Kjanichwara}) can be extended to a more
general class of polynomial  superluminal operations acting on qubits,
qutrit and higher dimensional qudits, such as:
\begin{equation}
Q_2(\phi) =  \left(\begin{array}{ll}   1  &  e^{i\phi}  \\  0  &  0
\end{array}\right),\hspace{0.5cm}
Q_3(\phi_1,\phi_2)
= \left(\begin{array}{lll} 1 & e^{i\phi_1} & e^{i\phi_2} 
\\ 0 & 0 & 0 \\ 0 & 0 & 0
\end{array}\right), ~{\rm etc.}
\label{eq:Kjanichwara2}
\end{equation}

By  definition,  $Q=Q_2(0)$  and  $Q^\prime=Q_3(0,0)$.   To  see  that
$Q_2(\phi)$ is a polynomial operation, it suffices to show that it can
be  simulated  using  only   polynomial  amount  of  standard  quantum
mechanical resources, which  we do in the following  theorem. Let {\bf
  BQP}$_c$  denote  the  complexity  class  of problems  that  can  be
efficiently solved  on a standard  quantum computer that can  access a
constant gate. Then:
\begin{thm}
{\bf BQP}$_c = $ {\bf BQP}.
\label{thm:bqpc}
\end{thm}

{\bf  Proof.}  It  is  clear that  any  problem in  {\bf  BQP} can  be
efficiently  solved using resources  of {\bf  BQP}$_c$, by  simply not
using the constant gates. Now let us consider the simulation the other
way.    Given  an  arbitrary   $(n+1)$-qubit  state   $|\psi\rangle  =
|\alpha\rangle|0\rangle + |\beta\rangle|1\rangle$, where the $n$-qubit
states  $|\alpha\rangle$ and  $|\beta\rangle$ are  neither necessarily
mutually  orthogonal nor normalized  and with  $|||\alpha\rangle||^2 +
|||\beta\rangle||^2 = 1$, the action  of $Q_2(\phi)$ on the last qubit
is    to     produce    $Q_2|\psi\rangle    =     (|\alpha\rangle    +
e^{i\phi}|\beta\rangle)|0\rangle \equiv |\psi^\prime\rangle|0\rangle$,
which      is      interpreted      as      ${\cal      N}      \equiv
\langle\psi^\prime|\psi^\prime\rangle          =          1          +
2\left[\cos(\phi)\Re(\langle           \alpha|\beta\rangle)          -
  \sin(\phi)\Im(\langle  \alpha|\beta\rangle)\right]$  copies  of  the
normalized      state      $|\underline{\psi}^\prime\rangle     \equiv
|\psi^\prime\rangle/\sqrt{\cal  N}$  and a  copy  of $|0\rangle$.   If
$|\alpha\rangle$ and $|\beta\rangle$ are mutually orthogonal, and thus
the reduced  density operator  for the last  qubit is diagonal  in the
computational  basis,   then  ${\cal   N}=1$,  and  no   such  special
interpretation is needed.

To  simulate  the production  of  $|\psi^\prime\rangle$ with  standard
quantum    resources,    one    first    applies    a    phase    gate
$\left( \begin{array}{cc}  1 &  0 \\ 0  & e^{i\phi}\end{array}\right)$
followed  by  a  Hadamard on  the  last  qubit,  to obtain  the  state
$(1/\sqrt{2})(|\psi^\prime\rangle|0\rangle                            +
|\psi^{\prime\prime}\rangle|1\rangle)$                            where
$|\psi^{\prime\prime}\rangle         =         |\alpha\rangle        -
e^{i\phi}|\beta\rangle$.   Measurement  on   the  last  qubit  in  the
computational     basis      yields     $|0\rangle$,     and     hence
$|\underline{\psi}^\prime\rangle$   in   the   first  register,   with
probability  $|||\psi^\prime\rangle/\sqrt{2}||^2 = {\cal  N}/2$, which
is  to say  that the  simulation  of $Q_2$  succeeds with  probability
$1/2$,  irrespective  of  $n$.   Similar  arguments  hold  for  $Q_3$,
etc.  Therefore,  the  class  of problems  efficiently  solvable  with
standard quantum  computation augmented by the  non-standard family of
constant gates is in {\bf BQP}. \hfill $\blacksquare$
\bigskip


It is worth noting that the  constant gate is quite different from the
following two  operations that appear to  be similar, but  are in fact
quite distinct.  The first operation is a  standard quantum mechanical
CP  map, polynomial and not  superluminal; the second
is exponential and consequently superluminal.

(a)  To  begin  with,  a  constant  gate  is  not  a  quantum  deleter
\cite{sri07},  in which  a  qubit  is subjected  to  a {\em  complete}
operation,  in  specific,  a  contractive  CP  map  that  prepares  it
asymptotically in a fixed state  $|0\rangle$.  The action of a quantum
deleter is  given by an  amplitude damping channel  \cite{nc00}, which
has an operator sum representation, respectively
\begin{equation}
\label{eq:qdele}
\rho_2 \longrightarrow  \sum_j E_j \rho_2  E_j^{\dag};\hspace{0.5cm}
\rho_3 \longrightarrow \sum_j E_j^{\prime} \rho_3 E_j^{\prime\dag},
\end{equation}
in  the qubit  case or when extended to  the qutrit case, 
with  the Kraus operators  given by Eq.
(\ref{eq:sriKjanichwaraA}) or (\ref{eq:sriKjanichwaraB}), respectively
\begin{subequations}
\label{eq:sriKjanichwara}
\begin{eqnarray}
E_1 &\equiv& \left(\begin{array}{ll} 1 & 0 \\ 0 & 0
\end{array}\right),~~ E_2 \equiv \left(\begin{array}{ll} 0 & 1 \\ 0 & 0 
\end{array}\right), \label{eq:sriKjanichwaraA} \\ E_1^{\prime} &\equiv&
\left(\begin{array}{lll} 1 & 0 & 0 \\ 0 & 0 & 0 \\ 0 & 0 & 0
\end{array}\right),~~ E_2^{\prime} \equiv \left(\begin{array}{lll} 
0 & 1 & 0 \\ 0 & 0 & 0 \\ 0 & 0 & 0 \end{array}\right),~~ E_3^{\prime} 
\equiv \left(\begin{array}{lll} 0 & 0 & 1 \\ 0 & 0 & 0 \\ 0 & 0 & 0 
\end{array}\right).
\label{eq:sriKjanichwaraB}
\end{eqnarray}
\end{subequations}
Unlike in the case of  $Q$, $Q^\prime$ or $Q^{\prime\prime}$, there is
no actual destruction of quantum information, but its transfer through
dissipative  decoherence  into  correlations with  a  zero-temperature
environment.  The  reduced density operator of  Bob's entangled system
remains  unaffected by Alice's  application of  this operation  on her
system.   The deleting action,  though nonlinear  at the  state vector
level, nevertheless acts linearly on the density operator.
  

(b) Next  we note that the  constant gate is quite  different from the
`post-selection'  operation,  which is  a  {\it deterministic}  rank-1
projection \cite{insel}.   Verbally, if the  constant gate corresponds
to  the   operation  ``for  all   input  states  $|j\rangle$   in  the
computational   basis,  set   the  output   state   to  $|\xi\rangle$,
independently of $j$, except  for a global phase", where $|\xi\rangle$
is  some fixed state,  then post-selection  corresponds to  the action
``for  all input  states $|j\rangle$,  if  $j \ne  \xi$, then  discard
branch $|j\rangle$".  Post-selective equivalents of $Q$ and $Q^\prime$
are
\begin{equation}
Q_{\bf PS} =  \left(\begin{array}{ll}   1  &  0  \\  0  &  0
\end{array}\right);\hspace{0.5cm}
Q ^{\prime}_{\rm PS} 
= \left(\begin{array}{lll} 1 & 0 & 0 \\ 0 & 0 & 0 \\ 0 & 0 & 0
\end{array}\right),
\label{eq:KjanichwaraS}
\end{equation}
{\it followed by renormalization}.   In particular, whereas the action
of   $Q$   on   the   first    of   two   particles   in   the   state
$(1/\sqrt{2})(|00\rangle + |11\rangle)$  leaves the second particle in
the state $(1/\sqrt{2})(|0\rangle +  |1\rangle)$, that of $Q_{\rm PS}$
leaves  the   second  particle  in  the  state   $|0\rangle$.   It  is
straightforward   to  see  that   post-selection  is   an  exponential
operation: acting it on the second qubit of $|\Psi_{\rm nc}\rangle$ in
Eq.  (\ref{eq:sat}),  and post-selecting on 1, we  obtain the solution
to SAT in one time-step.

The seemingly immediate conclusion  due to the fact $C^< \not\subseteq
E$  is  that  the WNHE  assumption  is  not  strong enough  to  derive
no-signaling,  and  would  have  to be  supplemented  with  additional
assumption(s), possibly purely  physically motivated ones, prohibiting
the physical realization of polynomial superluminal gates.

An alternative, highly unconventional reading of the situation is that
WNHE  is a  fundamental principle  of  the physical  world, while  the
no-signaling  condition  is  in  fact  not  universal,  so  that  some
polynomial   superluminal    gates   may   actually    be   physically
realizable. Quite surprisingly,  we may be able to  offer some support
for this viewpoint.  We believe  that constant gates of above type can
be quantum  optically realized  when a photon  detection is made  at a
{\em path singularity}, defined as a  point in space where two or more
incoming paths converge and terminate.  In graph theoretic parlance, a
path singularity  is a  terminal node in  a directed graph,  of degree
greater than  1.  

We  describe  in Section  \ref{sec:ent}  an  experiment that  possibly
physically realizes $Q$.  In principle, a detector placed at the focus
of a  convex lens realizes such  a path singularity.   This is because
the geometry of the ray  optics associated with the lens requires rays
parallel to the  lens axis to converge to  the focus after refraction,
while  the   destructive  nature  of  photon   detection  implies  the
termination  of the  path.  Although  conceptually  and experimentally
simple, the high  degree of mode filtering or  spatial resolution that
the experiment requires will be the main challenge in implementing it.
Indeed, we  believe this is the  reason that such  gates have remained
undiscovered so far.

Our argument here has implicitly  assumed that ${\bf P} \ne {\bf NP}$.
If it  turns out that  ${\bf P} =  {\bf NP}$, then even  the obviously
non-physical operations such as $G$  or $R$ would be polynomial gates,
and  the  WNHE   assumption  would  not  be  able   to  exclude  them.
Nevertheless,  the question  of existence  and testability  of certain
superluminal gates, which is the main result of this work, would still
remain valid  and of interest.   If polynomial superluminal  gates are
indeed found to exist (and  given that other superluminal gates do not
seem  to exist  anyway), this  would give  us greater  confidence that
${\bf P}  \ne {\bf  NP}$ (or, to  be safe,  that even Nature  does not
`know' that ${\bf P} = {\bf NP}$!)  and that the assumption of WNHE is
indeed a valid and fruitful one.

\section{Part II: Quantum optical test}

\subsection{An experiment with entangled pairs of photons \label{sec:ent}}

\begin{figure}
\includegraphics[width=17.0cm]{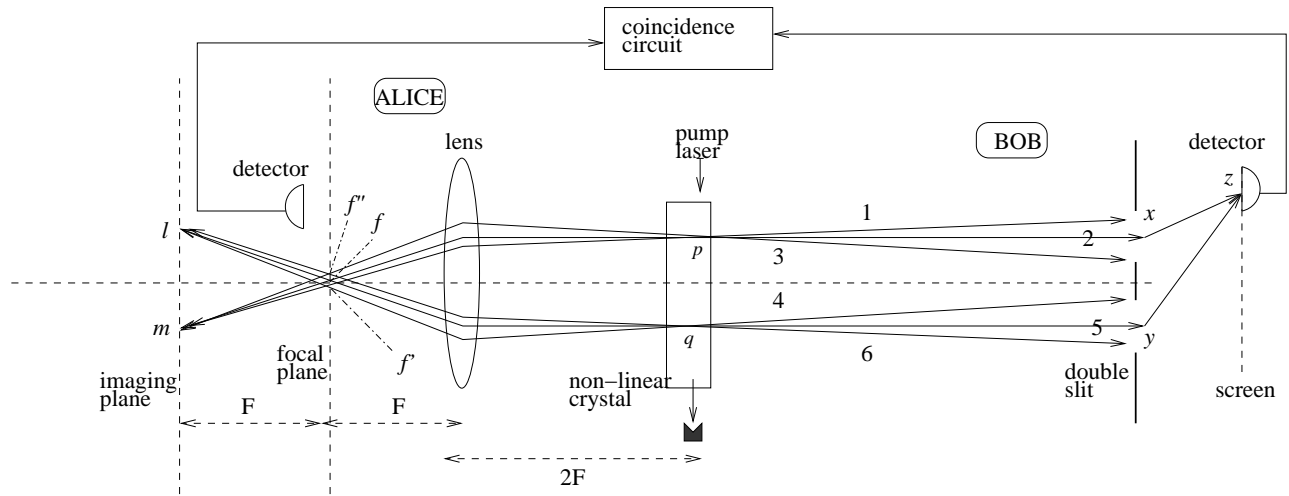} 
\caption{An  `unfolded' version  of the  Innsbruck experiment  (not to
  scale).  A  pair of momentum-entangled photons is  created by type-I
  parametric down  conversion of the pump laser.   Alice's photon (the
  signal  photon) is registered  by a  detector behind  a 
  lens.   Bob's photon (the  idler) is  detected behind  a double-slit
  assembly.  If her detector is placed in the focal plane
  of the  lens (of focal length  $F$), it projects Bob's  state into a
  mixture  of plane waves,  which produce  an interference  pattern on
  Bob's screen {\em in coincidence}  with any fixed detection point on
  Alice's focal plane. Bob's pattern in his {\em singles count}, being
  the integration of such patterns  over all focal plane points, shows
  no  interference  pattern.   On  the  other  hand,  positioning  her
  detector in the  imaging plane can potentially reveal the
  path the  idler takes through the  slit assembly, and  thus does not
  lead  to  an  interference  pattern  on Bob's  screen  even  in  the
  coincidence counts.}
\label{fig:zeiz}
\end{figure}

Our proposed implemention  of $Q^\prime$, based  on the  use of
entanglement,  is  broadly related  to  the  type  of quantum  optical
experiments encountered in Refs.   \cite{asp8}, and closely related to
an  experiment  performed   in  Innbruck  that  elegantly  illustrates
wave-particle duality by  means of entangled light \cite{zei00,dop98}.
In  the  Innsbruck experiment,  pairs  of position-momentum  entangled
photons  are  produced  by  means  of  type-I  spontaneous  parametric
down-conversion (SPDC) at  a nonlinear source, such as  a BBO crystal.
The two outgoing conical beams from the nonlinear source are presented
`unfolded'  in Figure \ref{fig:zeiz}.   One of  each pair,  called the
`signal photon',  is received  by Alice, while  the other,  called the
`idler',  is  received  and   analyzed  by  Bob.   Alice's  photon  is
registered by a detector behind a lens.

Bob's photon is  detected after it enters a  double-slit assembly.  If
Alice's detector, which  is located behind the lens,  is positioned at
the focal plane of the lens and detects a photon, it localizes Alice's
photon to a point on the focal plane.  By virtue of entanglement, this
projects the state of Bob's photon to a `momentum eigenstate', a plane
wave  propagating in a  particular direction.   For example,  if Alice
detects  her photon  at $f$,  $f^\prime$ or  $f^{\prime\prime}$, Bob's
photon is projected to a superposition  of the parallel modes 2 and 5,
modes 1 and 4, or modes  3 and 6.  Since this cannot reveal positional
information about whether  the particle originated at $p$  or $q$, and
hence reveals no which-way  information about slit passage, therefore,
{\em  in coincidence} with  a registration  of her  photon at  a focal
plane  point, the  idler exhibits  a Young's  double-slit interference
pattern  \cite{dop98,zei00}.  The  patterns  corresponding to  Alice's
registering her  photon at $f$, $f^\prime$  or $f^{\prime\prime}$ will
be  mutually shifted.   Bob's observation  in his  single  counts will
therefore not show any sign  of interference, being the average of all
possible such mutually shifted  patterns.  The interference pattern is
seen by Bob in coincidence  with Alice's detection.

If her detector is placed at the imaging plane (at distance
$2F$ from the  plane of the lens), a click of  the detector can reveal
the path  the idler takes from  the crystal through  the slit assembly
which  therefore cannot  show  the interference  pattern  even in  the
coincidence counts.  For  example, if Alice detects her  photon at $l$
(resp.,  $m$), Bob's  photon is  projected to  a superposition  of the
mutually  non-parallel modes  4, 5  and  6 (resp.,  1, 2  and 3)  and,
because the  double-slit assembly is  situated in the near  field, can
then  enter only slit  $y$ (resp.,  $x$).  Therefore,  Alice's imaging
plane  measurement gives  path or  position information  of  the idler
photon, so  that no interference pattern emerges  in Bob's coincidence
counts  \cite{dop98,zei00},  and  consequently  also  in  his  singles
counts. This  qualitative description  of the Innsbruck  experiment is
made  quantitative   using  a  simple  six-mode  model   in  the  next
Subsection.

\subsubsection{Quantum optical description of the Innsbruck experiment 
\label{sec:sridom}}

\begin{figure}
\includegraphics[width=17.0cm]{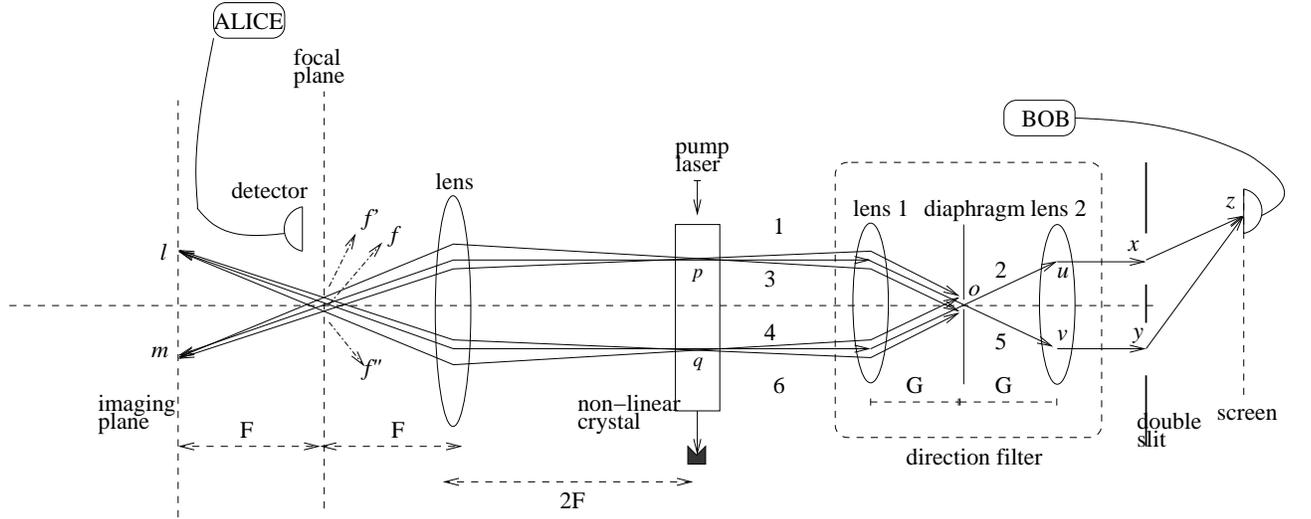} 
\caption{The  modified  Innsbruck  experiment  (not  to  scale):  Same
  configuration as in Figure  \ref{fig:zeiz}, except that Bob's photon
  (the idler),  before entering the double-slit  assembly, traverses a
  direction filter that permits only (nearly) horizontal modes to pass
  through,  absorbing  the  other  modes  at the  filter  walls.   The
  direction  filter acts  as  a  state filter  that  ensures that  Bob
  receives  only the  {\em pure}  state consisting  of  the horizontal
  modes.  Thus  if Alice makes no  measuement or makes  a detection at
  $f$, Bob's  corresponding photon  builds an interference  pattern of
  the modes 2 and 5 in the {\em singles} counts. On the other hand, if
  Alice positions  her detector  in the imaging  plane, she
  knows the  path the idler takes  through the slit  assembly. Thus no
  interference  pattern   is  found  on  Bob's  screen   even  in  the
  coincidence counts.}
\label{fig:zei}
\end{figure}

Here the
state of the SPDC field of Figure \ref{fig:zei} is modeled by a 6-mode
vector:
\begin{equation}
\label{spdc}
|\Psi\rangle =  (1 +  \frac{\epsilon}{\sqrt{6}}
\sum_{j=1}^6 a^{\dag}_jb^{\dag}_j)|{\rm vac}\rangle 
\end{equation}
where $|{\rm  vac}\rangle$ is  the vacuum state,  $a^{\dag}_j$ (resp.,
$b^{\dag}_j$) are  the creation  operators for Alice's  (resp., Bob's)
light  field on  mode $j$,  per the  mode numbering  scheme  in Figure
\ref{fig:zei}. The  quantity  $\epsilon$ ($\ll  1$)
depends on the pump field  strength and the crystal nonlinearity.  The
coincidence counting rate for  corresponding measurements by Alice and
Bob  is  proportional  to the  square  of  the  second-order  correlation
function, and given by:
\begin{equation}
\label{Eq:coinc}
R_\alpha(z)      \propto      \langle\Psi|E^{(-)}_z     E^{(-)}_\alpha
E^{(+)}_\alpha  E^{(+)}_z|\Psi\rangle = ||E^{(+)}_\alpha
E^{(+)}_z|\Psi\rangle||^2,   ~~~~   (\alpha   =   f,f^{\prime\prime},l,
m,\cdots).
\end{equation}
where $E_\alpha^{(+)}$  represents the positive frequency  part of the
electric  field at  a point  on Alice's  focal or  imaging  plane, and
$E_z^{(+)}$ that  of the electric field  at an arbitrary  point $z$ on
Bob's  screen.  We have:
\begin{equation}
\label{bobfjeld}
E_z^{(+)} = 
e^{ikr_D}\left(e^{ikr_2}\hat{b}_2 + e^{ikr_5}\hat{b}_5\right) +
e^{ikr_{D^\prime}}\left(e^{ikr_1}\hat{b}_1 + e^{ikr_4}\hat{b}_4\right) +
e^{ikr_{D^{\prime\prime}}}\left(e^{ikr_3}\hat{b}_3 + e^{ikr_6}\hat{b}_6\right), 
\end{equation}
where $k$ is the wavenumber, $r_D$ the distance from the EPR source to
the upper/lower slit on Bob's double slit diaphragm (the length of the
segment $\overline{qy}$  or $\overline{px}$); $r_2$  (resp., $r_5$) is
the distance from the lower (resp., upper) slit to $z$.  The other two
terms in  Eq. (\ref{bobfjeld}),  pertaining to the  other two  pair of
modes,   are  obtained   analogously.    We  study   the  two   cases,
corresponding  to Alice making  a remote  position or  remote momentum
measurement on the idler photons.

{\em Case  1. Alice remotely  measures position (path) of  the idler.}
Suppose Alice positions her dectector at the imaging plane
and detects a photon at $l$ or $m$.
The corresponding field at her detector is
\begin{equation}
\label{alicefjeldb}
E_m^{(+)} =  e^{ik s_{m}}(\hat{a}_1+  \hat{a}_2  + \hat{a}_3); \hspace{0.5cm}  
E_l^{(+)} = e^{ik s_{l}}(\hat{a}_4 + \hat{a}_5 + \hat{a}_6),
\end{equation}
where $s_{m}$ (resp.,  $s_{l}$) is the path length  along any ray path
from the  source point  $p$ (resp., $q$)  through the lens  upto image
point $m$ (resp., $l$). By  Fermat's principle, all paths connecting a
given pair  of source  and image point  are equal.  Setting  $\alpha =
l,m$  in Eq.  (\ref{Eq:coinc}),  and substituting  Eqs.  (\ref{spdc}),
(\ref{bobfjeld}) and (\ref{alicefjeldb})  in Eq.  (\ref{Eq:coinc}), we
find the coincidence counting rate  for detections by Alice and Bob to
be
\begin{equation}
\label{Rg}
R_{m}(z)  \propto   \epsilon^2|e^{ik r_1} + e^{ik r_2} + e^{ik r_3}|^2;
\hspace{0.5cm}
R_{l}(z)  \propto   \epsilon^2|e^{ik r_4} + e^{ik r_5} + e^{ik r_6}|^2,
\end{equation}  
which is essentially a  single slit diffraction pattern formed behind,
respectively,  the upper and lower  slit.  The  intensity  pattern Bob
finds  on his  screen  in  the singles  count,  obtained by  averaging
$R_{\alpha}(z)$   over  $\alpha=l,m$,  is   thus  not   a  double-slit
interference pattern, but an incoherent mixture of the two single slit
patterns.  A similar  lack of interference pattern is  obtained by Bob
if Alice  makes no measurement. 

{\em  Case 2.   Alice remotely  measures momentum  (direction)  of the
  idler.}  Alice  positions her  dectector on the  focal plane  of the
lens.  If  she  detects  a photon  at  $f$, $f^\prime$  or
$f^{\prime\prime}$, the field at her detector is, respectively,
\begin{subequations}
\label{alicefjelda}
\begin{eqnarray}
E_f^{(+)}  &=&  e^{ikr_{2f}}\hat{a}_2  +  e^{ikr_{5f}}\hat{a}_5
= e^{ikr_{f}}(\hat{a}_2  +  \hat{a}_5),  \label{alicefjeldaa} \\
E_{f^\prime}^{(+)}  &=&  e^{ikr_{1f^\prime}}\hat{a}_1  +  e^{ikr_{4f^\prime}}\hat{a}_4
= e^{ikr_{1f^\prime}}(\hat{a}_1  +  e^{ik(r_{5f^\prime}-r_{1f^\prime})}\hat{a}_4),
\label{alicefjeldab} \\
E_{f^{\prime\prime}}^{(+)}  &=& e^{ikr_{3f^{\prime\prime}}}\hat{a}_3 +
e^{ikr_{6f^{\prime\prime}}}\hat{a}_6                                  =
e^{ikr_{3f^{\prime\prime}}}(\hat{a}_3                                 +
e^{ik(r_{6f^{\prime\prime}}-r_{3f^{\prime\prime}})}\hat{a}_6),
\label{alicefjeldac} 
\end{eqnarray}  
\end{subequations}
where $r_{2f}$ (resp., $r_{5f}$) is the distance from $p$ (resp., $q$)
along the path 2 (resp., 5) path through the lens upto point $f$.  The
distances along the two paths being identical, $r_{2f} = r_{5f} \equiv
r_f$.       The      distances     $r_{1f^\prime},      r_{4f^\prime},
r_{3f^{\prime\prime}}$   and   $r_{6f^{\prime\prime}}$   are   defined
analogously.   Substituting Eqs.   (\ref{spdc}),  (\ref{bobfjeld}) and
(\ref{alicefjelda}) in Eq.   (\ref{Eq:coinc}), we find the coincidence
counting rate is given by
\begin{subequations}
\label{ram}
\begin{eqnarray}
R_f(z)  &\propto& \epsilon^2\left[1 +  \cos(k\cdot[r_2 -  r_5])\right],
\label{rama} \\
R_{f^\prime}(z) &\propto& \epsilon^2\left[1 + \cos(k\cdot[r_1 - r_4] +
  \omega_{14})\right],
\label{ramb} \\
R_{f^{\prime\prime}}(z) &\propto&  \epsilon^2\left[1 + \cos(k\cdot[r_3
    - r_6] + \omega_{36})\right],
\label{ramc}
\end{eqnarray}    
\end{subequations}
where $\omega_{14} \equiv k(r_{4f}  - r_{1f})$ and $\omega_{36} \equiv
k(r_{6f}  -  r_{3f})$  are  fixed  for  a given  point  on  the  focal
plane.  Each equation  in  Eq. (\ref{ram})  represents a  conventional
Young's double slit pattern. Conditioned on Alice detecting photons at
$f$,  Bob  finds  the  pattern  $R_f(z)$,  and  similarly  for  points
$f^\prime$  and   $f^{\prime\prime}$.   In  his   singles  count,  Bob
perceives  no interference,  because  he is  left  with a  statistical
mixture  of  the  patterns (\ref{rama}),  (\ref{ramb}),  (\ref{ramc}),
etc.,  corresponding  to  {\em  all}  points on  Alice's  focal  plane
illuminated by the signal beam.


\subsubsection{The proposed experiment \label{sec:sriprop}}

The  experiment  proposed  here,  presented  earlier  by  us  in  Ref.
\cite{sri}, is  derived from  the Innsbruck experiment,  and therefore
called  `the  Modified  Innsbruck  experiment'.   It  was  claimed  to
manifest  superluminal signaling,  though it  was not  clear  what the
exact origin of the signaling was, and in particular, which assumption
that goes to proving the no-signaling theorem was being given up.  The
Modified Innsbruck  experiment is revisited  here in order  to clarify
this issue in  detail in the light of the  discussions of the previous
Sections.   This will  help  crystallize  what is,  and  what is  not,
responsible for the claimed signaling effect.  In Ref.  \cite{sri888},
we  studied  a  version  of  nonlocal communication  inspired  by  the
original   Einstein-Podolsky-Rosen   thought  experiment   \cite{epr}.
Recently, similar experiments, also based on the Innsbruck experiment,
have been independently proposed in Refs. \cite{sricram,sridom}.

First  we present  a qualitative  overview of  the  modified Innsbruck
experiment.   The  only   material  difference  between  the  original
Innsbruck experiment and the modified  version we propose here is that
the  latter contains a  `direction filter',  consisting of  two convex
lenses  of the  same focal  length  $G$, separated  by distance  $2G$.
Their shared focal plane is covered  by an opaque screen, with a small
aperture  $o$ of  diameter $\delta$  at their  shared focus.   We want
$\delta$ to be  small enough so that only  almost horizontal modes are
permitted by  the filter  to fall on  the double slit  diaphragm.  The
angular spread  (about the horizontal) of  the modes that  fall on the
aperture  is given  by $\Delta  \theta  = \delta/G$,  we require  that
$(\delta/G)  \sigma   \ll  \lambda$,   where  $\sigma$  is   the  slit
separation, to guarantee that only modes that are horizontal or almost
horizontal  are selected  to  pass through  the  direction filter,  to
produce  a  Young's double-slit  interference  pattern  on his  screen
plane.  On the  other hand, we don't want the aperture  to be so small
that it produces significant  diffraction, thus: $\delta \gg \lambda$.
Putting these conditions together, we must have
\begin{equation}
1 \ll \frac{\delta}{\lambda} \ll \frac{G}{\sigma}. \label{eq:sricond}
\end{equation}
The  ability  to satisfy  this  condition,  while  preferable, is  not
crucial.  If  it is  not satisfied strictly,  the predicted  signal is
weaker  but not entirely  suppressed. The  point is  clarified further
down.

If Alice  makes no measurement,  the idler remains entangled  with the
signal photon,  which renders incoherent the beams  coming through the
upper  and  lower  slits on  Bob's  side,  so  that  he will  find  no
interference  pattern on his  screen.  Similarly,  if she  detects her
photon in  the imaging plane, she  localizes Bob's photon  at his slit
plane, and so,  again, no interference pattern is  seen. Thus far, the
proposed experiment the same effect as the Innsbruck experiment.

On  the  other hand,  if  Alice  scans the  focal  plane  and makes  a
detection, she remotely measures Bob's corresponding photon's momentum
and erases its path  information, thereby (non-selectively) leaving it
as a mixture of plane waves incident on the direction filter.  However
only  the  fraction  that  makes  up the  pure  state  comprising  the
horizontal modes  passes through  the filter. Diffracting  through the
double-slit diaphragm, it produces  a Young's double slit interference
pattern on  Bob's screen.  Those  plane waves coincident  with Alice's
detecting her photon  away from focus $f$ are filtered  out and do not
reach  Bob's double slit  assembly.  It  follows that  an interference
pattern  will emerge in  Bob's {\em  singles counts},  coinciding with
Alice's detection  at $f$  or close to  $f$.  Thus Alice  can remotely
prepare inequivalent ensembles of  idlers, depending on whether or not
she measures momentum on her photon.  In principle, this constitutes a
superluminal signal.

Quantitatively,  the only  difference  between the  Innsbruck and  the
proposed experiment  is that Eq.  (\ref{bobfjeld}) is replaced  by an
expression containing  only horizontal modes.  As  an idealization (to
be relaxed below), assuming perfect filtering and low spreading of the
wavepacket at the aperature, we have:
\begin{equation}
\label{bobfjeld0}
E_z^{(+)} = 
e^{ikr_D}\left(e^{ikr_2}\hat{b}_2 + e^{ikr_5}\hat{b}_5\right),
\end{equation}
where $r_D$ now represents the distance from the EPR source to
the upper/lower slit on Bob's double slit diaphragm (the length of the
segment $\overline{qoux}$  or $\overline{povy}$); $r_2$  (resp., $r_5$) is
the distance from the upper (resp., lower) slit to $z$.  
{\em Detection of a signal photon  at or near $f$ is the only possible
  event on  the focal plane such  that Bob detects the  twin photon at
  all}.   Focal plane  detections sufficiently  distant from  $f$ will
project the idler into non-horizontal  modes that will be filtered out
before   reaching   Bob's   double-slit  assembly.    Therefore,   the
interference pattern Eq.  (\ref{rama}) is in fact the only one seen in
Bob's singles counts.  We denote  by $R^F(z)$, this pattern, which Bob
obtains  conditioned  on  Alice  measuring  in the  focal  plane.   By
contrast, in the Innsbruck experiment Bob in his singles counts sees a
statistical  mixture  of   the  patterns  (\ref{rama}),  (\ref{ramb}),
(\ref{ramc}), etc., corresponding to {\em all} points on Alice's focal
plane illuminated by the signal beam.

When  Alice  measures  in  the  imaging plane,  as  in  the  Innsbruck
experiment  Bob   finds  no   interference  pattern  in   his  singles
counts.  Setting   $\alpha  =  l,m$  in   Eq.   (\ref{Eq:coinc}),  and
substituting     Eqs.      (\ref{spdc}),     (\ref{bobfjeld0})     and
(\ref{alicefjeldb}) in Eq.   (\ref{Eq:coinc}), we find the coincidence
counting rate for detections by Alice and Bob to be
\begin{equation}
\label{Rg0}
R_{\alpha}(z)  \propto   \epsilon^2,~~~~  (\alpha  =  l,   m),
\end{equation}  
which is a uniform pattern (apart  from an envelope due to single slit
diffraction, which we  ignore for the sake of  simplicity). It follows
that Bob's observed pattern in the singles counts conditioned on Alice
measuring  in the  imaging plane,  $R^I(z)$, is  also the  same, i.e.,
$R^I(z) \propto \epsilon^2$.

Our main  result is the  difference between the patterns  $R^I(z)$ and
$R^F(z)$,  which  implies  that  Alice  can  signal  Bob  one  bit  of
information across the spacelike interval connecting their measurement
events, by choosing to measure her photon in the focal plane or not to
measure.  In practice, Bob  would need to include additional detectors
to  sample or  scan the  $z$-plane  fast enough.   This procedure  can
potentially  form  the basis  for  a  superluminal quantum  telegraph,
bringing into sharp focus  the tension between quantum nonlocality and
special relativity.

Considering the far-reaching implications  of a positive result to the
experiment, we  may pause to consider  the following:
whether our analysis  of so far
can be correct, and-- under the possibility 
(however limited) that it is-- how
such a signal may ever arise, in view of the no-signaling theorem.  It
may be easy to dismiss  a proof of putative superluminal communication
as `not even  wrong', yet less easy to spot  where the purported proof
fails and to provide a mechanism for thwarting the signaling. 
For  one, the prediction  of the nonlocal signaling  is based on a
model that  departs only slightly  from our quantum optical  model of
Section \ref{sec:sridom}, which  explains the original Innsbruck 
experiment quite
well. There have been various attempts at proving that quantum
nonlocality  somehow contravenes special  relativity.  The  author has
read some of their accounts, and it was not difficult to spot a hidden
erroneous assumption that led to the alleged conflict with relativity.
Armed with  this lesson,  the present claim  will be different  in the
following three ways:

\begin{itemize}
\item {\em We discuss in the following Section various
  possible objections to  our claim, and demonstrate why  each of them
  fails.} Perhaps they do not cover some erroneous but subtle assumption,
but even so, our present exercise could still
be instructive  in yielding new  theoretical insights. For  example, a
proposal for  superluminal communication based  on light amplification
was eventually  understood to fail because it  violates the no-cloning
theorem, a principle  that had not been discovered at  the time of the
proposal was made (cf. \cite{peres}).

\item {\em We single out, in the following Section, the key assumption
  responsible for  the superluminality} (that Alice's
  momentum measurement implements a polynomial superluminal gate). 
  This singling out of the non-standard element  at play makes it easier  for the
  reader to judge whether the  proposal is wrong, not even wrong, or--
  as we believe is the case-- worth testing experimentally.

\item  {\em We have  furnished computation-  and information-theoretic
  grounds  for why  superluminal gates  could be  possible,}   
according to which no-signaling  could  be a  nearly-universal-but-not-quite
side  effect of  the  computation theoretic  properties of  physical
reality; elsewhere \cite{ganeshchandru},  we show how the relativity
  principle    could   be   a    consequence   of    conservation   of
  information. 
\end{itemize}

In the  last Section, we clarify  how non-complete measurements, if experimentally
validated, could possibly  fit in with
known  physics.  There  we will  argue that  they arise  owing  to the
potential fact  that practically measurable  quantities resulting from
quantum  field theory  are not  described by  hermitian  operators, at
variance with a key axiom of orthodox quantum theory \cite{srihrv}.

\subsection{The question of existence and
origin of the signaling \label{sec:sriung}}

In the  Section, we will consider  a number of  possible objections to
our  main result, and demonstrate quantitatively why each of them
fails.

\paragraph{Spreading at the direction filter.} 
It can be shown that the effect of spreading at the direction filter 
only lowers-- but does  not eliminate-- the distinguishability between
the two  kinds of pattern  that Bob receives.  
For illustration,  suppose $\delta =  10\lambda$, and  as a
result, nearly only horizontal modes $r_2$ and $r_5$ are selected, but
the diffractive spreading at the filter  is strong, assumed to be given by
$\left(\begin{array}{cc}\cos\theta      &     \sin\theta
  \\ -\sin\theta & \cos\theta\end{array}\right)$ in the space spanned
by modes 2 and 5, where $\theta$ is determined by the
  geometry of  the filter.   In place of  Eqs. (\ref{bobfjeld}) 
and (\ref{Rg}), we now have:
\begin{eqnarray}
\label{eq:Bobfjeldnew}
E_z^{\prime  (+)}  &=&  e^{ikr_D} \left(e^{ikr_2}(\cos\theta\hat{b}_2  +
\sin\theta\hat{b}_5)  +  e^{ikr_5}(\cos\theta  \hat{b}_5 -  \sin\theta
\hat{b}_2) \right). \nonumber \\
R_\alpha^{\prime}(z)        &\propto&        \epsilon^2\left[1        \pm
\sin(2\theta)\cos(k\cdot[r_2 - r_5])\right], ~~~
({\rm with~}\pm {\rm ~according~as~} \alpha=l,m).
\end{eqnarray}
The pattern found by Bob  in  his singles  counts  is  
$R_l^{\prime}(z)  + R_m^{\prime}(z) \propto \epsilon^2$,
which is  a constant  pattern (ignoring  the finite
width of the slits), just as when the spreading had
been ignored.  On the  other hand, in place of Eq. (\ref{ram}), we now obtain
\begin{equation}
\label{sriram}
R^{\prime}_f(z)   \propto   
\epsilon^2\left[1  +   \cos(2\theta)\cos(k\cdot[r_2   -  r_5])\right],
\end{equation}    
Except  in   the  case
$\theta=\pi/4$,  which is  highly unlikely,  and in  any case,  can be
precluded by altering $\delta$ or $G$, the two patterns are in principle
distinguishable.

\paragraph{Alice's focal plane measurement implements a 
constant gate in the subspace of interest}

The  state (\ref{spdc})  is now  represented in  a simple  way  as the
unnormalized state
$
|\Psi^{(1)}\rangle           =
\frac{\epsilon}{\sqrt{6}}\sum_{j=1}^6|j,j\rangle,
$
where for  simplicity the vacuum  state, which does not  contribute to
the entanglement related  effects, is omitted, and it  is assumed that
each mode  contains at  most one pair  of entangled photons  (i.e., no
higher  excitations  of the  light  field).   Further  because of  the
direction filter, it suffices to restrict our attention to the state
\begin{equation}
|\psi^{(2)}\rangle \propto \frac{1}{\sqrt{2}}(|2,2\rangle  + |5,5\rangle),
\label{eq:srispdc}
\end{equation}
the projection of $|\Psi^{(1)}\rangle$  onto ${\cal H}_2 \otimes {\cal
  H}_2$, where  ${\cal H}_2$ is the subspace  spanned by $\{|2\rangle,
|5\rangle\}$.  Under  these assumptions, Alice's  position measurement
in  this  subspace,  represented  by  the  operators  $\hat{a}_2$  and
$\hat{a}_5$, can  be written as the Kraus  operators $\hat{a}_2 \equiv
|0\rangle\langle  2|$  and $\hat{a}_5  \equiv  |0\rangle \langle  5|$.
Within  ${\cal  H}_{2}$ these  operators  form  a  complete set  since
$\hat{a}^{\dag}_2\hat{a}_2 + \hat{a}^{\dag}_5\hat{a}_5 =
|2\rangle\langle  2|  + |5\rangle\langle  5|  = \mathbb{I}_2$.   Thus,
Alice's measurement on $|\Psi^{(2)}\rangle$ in the position basis does
not nonlocally affect Bob's reduced density operator,
which is proportional to $\mathbb{I}_2/2$.

On  the other  hand, if  Alice  measures momentum,  her measurment  is
represented    by   the    field   operator    $E^{(+)}_f$    in   Eq.
(\ref{alicefjelda}). We have in the above notation
\begin{equation}
E_f^{(+)} \propto \hat{a}_2 + \hat{a}_5
\equiv |0\rangle(\langle 2| + \langle 5|).
\label{eq:srif0}
\end{equation}
This is  just the polynomial  superluminal gate $Q$  in Eq.
(\ref{eq:sriKjanichwara}), with the output basis given by
$\{|0\rangle, |0^\perp\rangle\}$, where $|0^\perp\rangle$ is
any basis element orthogonal to the vacuum state.

We note that the operator
$E_{\overline{f}}^{(+)} \propto \hat{a}_2 - \hat{a}_5
\equiv |0\rangle(\langle 2| - \langle 5|)$, that would complete
$E_f^{(+)}$ in that 
$E_f^{(-)}E_f^{(+)} +
E_{\overline{f}}^{(-)}E_{\overline{f}}^{(+)} = \mathbb{I}$ in the
space span$(|2\rangle, |5\rangle)$. However,  
$E_{\overline{f}}^{(+)}$ is necessarily non-physical in the given
geometry since modes 2 and 5 meets only at $f$, where the
electric field operator is indeed
$\propto \hat{a}_2 - \hat{a}_5$.

We further note that, inspite of the non-completeness of $E_f^{(+)}$, the
structure of $|\psi^{(2)}\rangle$ in Eq. (\ref{eq:srispdc})
guarantees that $E_f^{(+)}|\psi^{(2)}\rangle$ is by default normalized,
and hence poses no problem with respect to probability conservation.

By  contrast,  Bob's  measurement  is  complete  (which  rules  out  a
Bob-to-Alice  superluminal signaling).  Each  element of  Bob's screen
$z$-basis  is  a  possible  outcome,  described  by  the  annihilation
operator approximately of  the form $\hat{E}^{(-)}_z \propto \hat{a}_2
+  e^{i\gamma}\hat{a}_5$, where  $\gamma =  \gamma(k,z)$ is  the phase
difference between the paths 2 and 5  from the slits to a point $z$ on
Bob's    screen.    This    represents    a   POVM    of   the    form
$\hat{E}^{(-)}_z\hat{E}^{(+)}_z =  (|2\rangle + e^{-i\gamma}|5\rangle)
(\langle 2| +  e^{i\gamma}\langle 5|)$.  Even though $\hat{E}^{(+)}_z$
has the same  form as Alice's operator $\hat{E}^{(+)}_f$--  as a Kraus
operator describing the absorption of two interfering modes at a point
$z$--,  yet, when integrated  over his  whole `position  basis', Bob's
measurement is seen  to form a complete set, for, as  it can be shown,
$\int_{z=-\infty}^{+\infty}    \hat{E}^{(-)}_z\hat{E}^{(+)}_z   dz   =
|2\rangle\langle  2| + |5\rangle\langle  5|$. In  the case  of Alice's
momentum  measurement,  because  the   detection  happens  at  a  path
singularity, a  similar elimination of cross-terms  via integration is
not  possible, whence  the  non-completeness.  It  is indeed  somewhat
intriguing how  geometry plays a  fundamental role in  determining the
completeness status  of a  measurement. This has  to do with  the fact
that the direct  detection of a photon is  practically a determination
of  {\em  position}  distribution.   For  example,  even  in  remotely
measuring the  idler's momentum, Alice measures  her photon's position
at the focal  plane.  We will return again to this  issue in the final
Section.


\paragraph{Role of the direction filter.} 
A simple model of the action of the perfect direction filter is
\begin{equation}
D  \equiv  \sum_{j=2,5} |j\rangle\langle  j|  + \sum_{j\ne  2,5}
|{\rm-}j\rangle\langle j|
\end{equation}
acting   locally   on   the   second   register  of   the   state   of
$|\Psi^{(1)}\rangle$.           =
Here  $|{\rm -}j\rangle$ can be thought  of as a
state orthogonal to all  $|j\rangle$'s and other $|{\rm -}j\rangle$'s,
that  removes  the  photon   from  the  experiment,  for  example,  by
reflecting it out or by absorption  at the filter. It suffices for our
purpose  to  note that  $D$  can be  described  as  a local,  standard
(linear, unitary and hence complete) operation. Since the structure of
QM  guarantees  that  such   an  operation  cannot  lead  to  nonlocal
signaling,  the conclusion  is  that the  superluminal  signal, if  it
exists, must remain {\em even if the the direction filter is absent}.

We     will    employ     the    notation     $|j+k+m\rangle    \equiv
(1/\sqrt{3})(|j\rangle  + |k\rangle  + |m\rangle)$.   To see  that the
nonlocal  signaling  is implicit  in  the  state  modified by  Alice's
actions  even without  the  application  of the  filter,  we note  the
following:    if   Alice    measures   `momentum'    on    the   state
$|\Psi^{(1)}\rangle$ and detects a  signal photon at $f$, she projects
the corresponding idler into  the state $|2+5\rangle$.  Similarly, her
detection of  a photon at  $f^{\prime\prime}$ projects the  idler into
the state  $|3+6\rangle$, and her detection  at $f^{\prime}$, projects
the idler into the state  $|1+4\rangle$.  Therefore, in the absence of
the  direction  filter,  Alice's  remote measurement  of  the  idler's
momentum  leaves  the idler  in  a  (assumed  uniform for  simplicity)
mixture given by
\begin{equation}
\rho_P \propto |2+5\rangle\langle2+5| + |1+4\rangle\langle1+4|
+ |3+6\rangle\langle3+6|.
\label{eq:srirhop}
\end{equation}
Her momentum measurement is non-complete, since the summation over the
corresponding projectors (r.h.s of  Eq. (\ref{eq:srirhop})) is not the
identity  operation  $\mathbb{I}_6$ pertaining  to  the Hilbert  space
spanned by six modes $|j\rangle$ $(j=1,\cdots,6)$.

On the  other hand, if  Alice remotely measures the  idler's position,
she leaves the idler in the mixture
\begin{equation}
\rho_Q \propto |1+2+3\rangle\langle1+2+3| + |4+5+6\rangle\langle4+5+6|.
\label{eq:srirhoq}
\end{equation}
Here again, her position measurement is non-complete, reflected in the
fact that  the summation over  the corresponding projectors  (r.h.s of
Eq. (\ref{eq:srirhoq})) is not $\mathbb{I}_6$ \cite{sridir}.

Since $\rho_P \ne  \rho_Q$, we are led to  conclude that the violation
of no-signaling {\em is already implicit in the Innsbruck experiment}.
Yet, since Bob measures in the $z$-basis rather than the `mode' basis,
in the absence of a direction filter-- as is the case in the Innsbruck
experiment--,  Bob's screen  will  not register  any  signal, for  the
following  reason.  In case  of Alice's  focal plane  measurement, the
integrated diffraction-interference pattern corresponding to different
outcomes will  wash out any  observable interference pattern.   On the
other hand, in the case of Alice's imaging plane  measurement, each of
Bob's
detections comes  from the photon's  incoherent passage through  one or
the  other  slit,  and  hence--  again-- no  interference  pattern  is
produced on his  screen.  Thus, measurement at Bob's  screen plane $z$
without   a  direction   filter  will   render   $\rho_P$  effectively
indistinguishable  from $\rho_Q$.   The role  played by  the direction
filter  is to  prevent modal  averaging  in case  of Alice's  momentum
measurement,  by selecting  one set  of  modes.  The  filter does  not
create, but only exposes, a superluminal effect that otherwise remains
hidden.


\paragraph{Complementarity of single- and two-particle correlations.}

It  is well  known  that  path information  (or  particle nature)  and
interference (or wave nature) are mutually exclusive or complementary.
In the two-photon case, this  takes the form of mutual incompatibility
of single-  and two-particle interference  \cite{abu01,bos02}, because
entanglement  can be  used to  monitor  path information  of the  twin
particle, and is  thus equivalent to `particle nature'.   One may thus
consider single-  and two-particle correlations as being  related by a
kind   of   complementarity   relation   that  parallels   wave-   and
particle-nature complementarity.  A brief exposition of this idea
is given in the following paragraph.

For a particle in a  double-slit experiment, we restrict our attention
to the Hilbert space ${\cal  H}$, spanned by the state $|0\rangle$ and
$|1\rangle$ corresponding to the upper and lower slit of a double slit
experiment.  Given density operator $\rho$, we define coherence $C$ by
$C =  2|\rho_{01}| = 2 |\rho_{10}|$,  a measure of  cross-terms in the
computational basis  not vanishing. The particle  is initially assumed
to be  in the state  $|\psi_a\rangle$, and a ``monitor",  initially in
the  state $|0\rangle$,  interacting with  each other  by means  of an
interaction $U$, parametrized by variable $\theta$ that determines the
entangling  strength  of  $U$.   It  is  convenient  to  choose  $U  =
\cos\theta~I\otimes I  + i\sin\theta {\rm  ~CNOT}$, where CNOT  is the
operation $I\otimes|0\rangle\langle0| + X \otimes |1\rangle\langle1|$,
where $X$  is the  Pauli $X$  operator. Under the  action of  $U$, the
system    particle    goes    to    the    state    $\rho    =    {\rm
  Tr}_m[U(|\psi_a\rangle|0\rangle\langle\psi_a|   \langle0|)U^\dag]  =
\frac{I}{2}   +  \frac{1}{2}[  (\cos\theta   +  i\sin\theta)\cos\theta
  |0\rangle\langle1|  +  {\rm  c.c}$, where  Tr$_m[\cdots]$  indicates
  taking  trace over  the  monitor.  Applying  the  above formula  for
  coherence to  $\rho$, we calculate that coherence  $C = \cos\theta$.
  We   let   $\lambda_{\pm}$  denote   the   eigenvalues  of   $\rho$.
  Quantifying the degree of entanglement by concurrence \cite{woo}, we
  have  $E \equiv 2\sqrt{\lambda_-\lambda_+}  = \sin\theta$.   We thus
  obtain a trade-off between  coherence and entanglement given by $C^2
  +  E^2  =  1$,   a  manifestation  of  the  complementarity  between
  single-particle and two-particle interference.

In  the context  of  the  proposed experiment,  this  could raise  the
following purported objection to our proposed signaling scheme: as the
experiment  happens  in  the  near-field  regime,  where  two-particle
correlations are strong, one would  expect that Bob should not find an
interference  pattern in his  singles counts.   Yet, contrary  to this
expectation,  Eq.   (\ref{ram})  implies  that  such  an  interference
pattern  does  appear.   The  reason   is  that  in  the  focal  plane
measurement,  Alice is  able  to  erase her  path  information in  the
subspace   ${\cal   H}_2$,   but,   by  virtue   of   the   associated
non-completeness, she  does so  in {\em only  one} way, viz.   via the
non-complete operation $E_f^{(+)}$ associated with her measurment.  If
her measurement were {\em  complete}, she would erase path information
in   more   than   one   way,  and   the   corresponding   conditional
single-particle interference patterns would mutually cancel each other
in the singles count. This is clarified in the following Section.
 

\paragraph{Polarization and `interferometric quantum computing'.}

$Q$-like gates  describe the situation  where two converging  modes at
the path  singularity have the  {\em same} polarization.   The quantum
optics formalism implies that if the polarizations of the two incoming
modes are not parallel  when interfering, then the polarization states
add  vectorially (that  is,  superpose), with  amplitudes being  added
componentwise  along each  polarization/dimension,  and the  resulting
probability being the  squared magnitude of this vector  sum.  One can
define a  corresponding more  general constant gate  (a tensor  sum of
constant gates  over the  internal dimensions), and  a correspondingly
potentially  larger  {\bf BQP}$_c$.   It  can  be  shown that  Theorem
\ref{thm:bqpc}  still  holds.   Here  we  will  content  ourselves  to
illustrate it by a simple example.

Suppose we have this `interferometric quantum computer': a $2^n$-level
atom, whose  spin part is  prepared initially in the  state $|a\rangle
\equiv (2^{-n/2})(|1\rangle  + |2\rangle + \cdots  + |2^n{-}1\rangle +
|2^n\rangle)$.   The spatial  part of  the atom's  matter wave  is now
split  into two  subwaves by  an appropriate  beam-splitter,  and then
refocused onto a path singularity.   On the second subwave, before the
two subwaves reach the region  of spatial overlap, an oracle operation
is applied  which in a single step  inverts the sign of  all the kets,
except  the `marked' state  $|2^n\rangle$, yielding  $|b\rangle \equiv
(2^{-n/2})(-|1\rangle   -|2\rangle   -\cdots   -   |2^n{-}1\rangle   +
|2^n\rangle)$.  According to the above prescription, the output at the
path    singularity    should   be    $|a\rangle    +   |b\rangle    =
2|2^n\rangle/2^{n/2}$,  i.e.,   a  particle  is   detected  with  with
exponentially   low  probability   $|||a\rangle   +  |b\rangle||^2   =
4\cdot2^{-n}$,  and  detection  leaves   the  particle  in  the  state
$|2^n\rangle$.   The  oracle  together  with  detection  at  the  path
singularity    is   equivalent    to   the    non-complete   operation
$\bigoplus_{j=1}^{2^n-1} Q^{(j)}_2(\pi) \oplus Q^{(2^n)}_2(0)$.

If the marked state is designated to be a possible solution to a SAT
problem, the measurement would  have to be
repeated  an exponentially large  number of  times, or performed once on
an exponentially large number of atoms, to  detect  a possible
`yes'  outcome.  Either  way, the  physical situation  is compatible
with the WNHE assumption, but not with no-signaling.
(We observe that augmenting the detection
with a  renormalization following  vector addition  
would in fact implement the post-selection gate.)  

Finally,  let  us  clarify  the sense  in  which  non-complete
operations  like   $Q$  of   potential  physical  interest   may  {\it
effectively}  conform to  probability  conservation.  
In  the Modified  Innsbruck experiment, Alice's application  of $Q$
conforms {\em exactly} to probability conservation, because 
the state  $|\psi^{(2)}\rangle$ in Eq. (\ref{eq:srispdc}) has a Schmidt
form, with $Q$ defined in Bob's Schmidt basis. However,  this
is not the general situation.  In such cases, one seems to
find that the spreading of the wavefunction produces
a pattern of bright and dark interferometric fringes at and around the
path singularity such that, even though locally there is an excess
or deficit  over the  average probability density,  still there  is an
overall probability conservation across the fringes.  This is somewhat
comparable      to       the  situation with       Bob's      POVM
$\hat{E}^{(-)}_z\hat{E}^{(+)}_z$,  which,
even though  locally a $Q$-like operation, still  yields identity when
integrated over $z$. This conservation 
mechanism is not applicable to the Modified Innsbruck experiment,
which is performed in the near-field limit, where spreading is minimal
and two-particle correlations are strong. However, as noted above, 
probability conservation is inherently exact for the situation 
in the experiment, and the mechanism need not be invoked.



As an illustration of the mechanism, let  the angle at which  the two interfering  beams of the
`interferometric quantum  compter' converge towards  a spatial overlap
region be  $\theta$.  The  fringes are given  by a  stationary pattern
with  spatial frequency $k^\prime  = k\sin\theta  \approx k\theta  = k
(S/d)$,  where  $S$ is  the  spatial  separation  between two  optical
elements (say,  mirrors) that are, respectively,  reflecting the beams
along the two  interferometric arms towards $q$, and  $d$ the distance
from the central  point between these mirrors to  the center of region
$q$.   The width  of each  fringe  is about  $2\pi/k^\prime =  \lambda
(d/S)$.  Now  the initial beam width  must be of the  order of several
wavelengths, and  the diffractive  spread rate of  each beam  at least
$\lambda/S$, so that beam width  $> \lambda d/S$.  Thus, the spreading
of (quantum)  waves guarantees that there will  always be compensatory
fringes, and  hence overall  conservation of probability,  even though
locally  the dark  and  bright bands  contain  less or  more than  the
average probability density.

Applied to the above atom interferometer example, 
the state vector at the interference screen will
have the form  $\kappa(\theta)(|a\rangle + e^{i\theta}|b\rangle)$ with
$\theta$ running  from $-\infty$ to  $+\infty$, where $\kappa(\theta)$
is  a  narrow Gaussian-like  function  centered  at $\theta=0$.   When
$\theta=0,  2\pi,  4\pi,\cdots$, one  obtains  dark  fringes with  the
`solution'  $|2^n\rangle$  at exponentially  low  intensity, as  noted
above.   When  $\theta=\pi,  3\pi,5\pi,  \cdots$, one  obtains  bright
fringes   of  nearly   maximal  intensity,   diminished  by   only  an
exponentially small  amount, corresponding, again,  to the `solution'.
Thus the  interference pattern is a band  of bright and  dark fringes at
spatial frequency $k^\prime$ with the bright ones very slightly dimmer
than if $|a\rangle$ and $|b\rangle$ had the same polarization, and the
dark ones very slightly brighter. 

\section{Discussions and Conclusions}\label{sec:conclu} 

Considering the far-reaching implications  of a positive result to our
proposed  experiment,  even  though  we  have  ruled  out  in  Section
\ref{sec:sriung} all the  (as far as we know) obvious objections,
we have to  remain open to the possibility that there  may be a subtle
error,  possibly a  hidden  unwarrented assumption,  somewhere in  our
analysis.  In the surprising event that the proposed experiment yields
a  positive outcome, 
no-signaling would no  longer be a universal condition,  and the issue
of  `speed  of   quantum  information'  \cite{srisal08}  would  assume
practical significance.  It would  also bolster the case for believing
that the WNHE assumption is  a basic principle of quantum physics, and
that    considerations   of    intractability,   and    by   extension
uncomputability, can serve as an informal guide to basic physics.

Physical  space  would be  regarded  as  a  type of  information,  and
physical dynamics  a kind of computation, with  physical separation 
being not
genuine obstacle  to rapid communication in  the way it  would be when
seen from  the perspective of  causality in conventional  physics.  On
the other hand, the  barrier between polynomial-time and hard problems
would  be real,  and the  physical existence  of  superluminal signals
would  thus  not  be  as  surprising as  that  of  exponential  gates.
Interestingly,  polynomial  superluminal   operations  exist  even  in
classical computation  theory.  The Random Access  Machine (RAM) model
\cite{ste09},  a standard  model  in computer  science wherein  memory
access  takes  exactly  one  time-step irrespective  of  the  physical
location of the memory element, illustrates this idea.  RAMs are known
to be polynomially equivalent to Turing machines.

Even granting  that the  noncomplete gate $Q^\prime$  turns out  to be
physically valid  and realizable, this brings us  to another important
issue: how  would non-completeness fit in with  the known mathematical
structure  of the  quantum  properties of  particles  and fields?   We
venture that the answer has to  do with the nature of and relationship
between  observables in  QM  on the  one  hand, and  those in  quantum
optics,  and more  generally, in  quantum field  theory (QFT),  on the
other hand.

It is frequently claimed that QFT  is just the standard rules of first
quantization  applied to classical  fields, but  this position  can be
criticized    \cite{srizeh,srihrv0,srihrv}.     For    example,    the
relativistic  effects   of  the   integer-spin  QFT  imply   that  the
wavefunctions describing a fixed number  of particles do not admit the
usual  probabilistic interpretation \cite{srihrv0}.   Again, fermionic
fields do not really have a classical counterpart and do not represent
quantum observables \cite{srihrv}.

In practice, measurable properties resulting from a QFT are properties
of particles--  of photons in quantum  optics.  Particulate properties
such  as number,  described by  the number  operator  constructed from
fields,  or the momentum  operator, which  allows the  reproduction of
single-particle QM in  momentum space, do not present  a problem.  The
problem is  the {\em position} variable,  which is considered  to be a
parameter,   and  not   a  Hermitian   operator,  both   in   QFT  and
single-particle relativistic QM,  and yet relevant experiments measure
particle positions.   The experiment described in  this work involves
measurement  of the  positions  of photons,  as  for example,  Alice's
detection of photons at points on the imaging or focal plane, or Bob's
detection at points on the $z$-plane, respectively.  There seems to be
no way to derive from  QFT the experimentally confirmed Born rule that
the nonrelativistic wavefunction  $\psi({\bf x},t)$ determines quantum
probabilities  $|\psi({\bf x},t)|^2$ of  particle positions.   In most
practical situations, this is really not a problem.  The probabilities
in the  above experiment were computed according  to standard quantum
optical rules to determine the correlation functions at various orders
\cite{glauber},  which  serve  as  an effective  wavefunction  of  the
photon,  as seen  for example  from Eqs.   (\ref{Eq:coinc}).   In QFT,
particle  physics phenomenologists have  developed intuitive  rules to
predict distributions of particle positions from scattering amplitudes
in {\em momentum} space.

Nevertheless, there  is a  problem in principle,  and this leads us  to ask
whether QFT is  a genuine quantum theory \cite{srihrv}.   If we accept
that properties like position are  valid observables in QM, the answer
seems to  be `no'.  We see this  again in the fact  that the effective
'momentum'  and  'position'  observables   that  arise  in  the  above
experiment are not seen to be Hermitian operators of standard QM (cf.
note   \cite{sridir}).     Further,   non-complete   operations   like
$\hat{E}^{(+)}_f$,  disallowed in  QM, seem  to appear  in  QFT.  This
suggests that  it is QM,  and not QFT,  that is proved to  be strictly
non-signaling by the no-signaling theorem.

Since nonrelativistic  QM and QFT  are presumably not  two independent
theories describing  entirely different  objects, but do  describe the
same   particles  in   many  situations,   the   relationship  between
observables  in  the  two  theories  needs to  be  better  understood.
Perhaps some  quantum mechanical observables are  a coarse-graining of
QFT  ones,  having wide  but  not  universal  validity.  For  example,
Alice's  detection of  a photon  at  a point  in the  focal plane  was
quantum mechanically  understood to project the state  of Bob's photon
into   a  one-dimensional   subspace  corresponding   to   a  momentum
eigenstate.    Quantum  optically,   however,  this   `eigenstate'  is
described as a  superposition of a number of  parallel, in-phase modes
originating  from different down-conversion  events in  the non-linear
crystal, producing  a coherent plane wave propagating  in a particular
direction.

\acknowledgments  I am  thankful to Prof.
S.   Rangwala   and  Mr.  K.   Ravi  for  insightful   elucidation  of
experimental issues, and to Prof.  J.  Sarfatti, Prof.  J.  Cramer and
Prof.  A. Shiekh for helpful discussions.


\begin{thebibliography}{100}
\bibitem{epr} A. Einstein, N. Rosen, and B. Podolsky.
Is the Quantum-Mechanical Description of Physical Reality Complete?
Phys. Rev. {\bf 47}, 777 (1935).
\bibitem{nos}  P.   H.  Eberhard,  Nuovo  Cimento   46B,  392  (1978);
C. D.  Cantrell and M.  O. Scully, Phys.  Rep.  {\bf 43},  499 (1978).
G. C. Ghirardi, A. Rimini, and T. Weber, Lett. Nuovo Cimento {\bf 27},
293  (1980).   P.  J.  Bussey,  Phys.  Lett.  {\bf  A  90},  9  (1982).
T. F.  Jordan, Phys. Lett. {\bf  94}A, 264 (1983).  A.  J. M. Garrett,
Found. Phys. {\bf  20} 381 (1990);  
A. Shimony, in {\em Proc. of
the Int. Symp. on Foundations  of Quantum Mech.}, eds. S. Kamefuchi et
al. (Phys.  Soc. Japan, Tokyo, 1993);   R Srikanth, Phys  Lett. {\bf A
292}, 161 (2002).
\bibitem{bell} J. S. Bell, Physics {\bf 1}, 195 (1964); J. F. Clauser,
M. A.  Horne, A. Shimony, and R.  A. Holt, Phys. Rev.  Lett. {\bf 23},
880 (1969); A.  Aspect, P.  Grangier and G.  Roger, Phys.  Rev.  Lett.
{\bf  49}, 91  (1982); W.   Tittel, J.   Brendel, H.   Zbinden  and N.
Gisin,  Phys.  Rev.   Lett  {\bf   81}  3563  (1998);  G.   Weihs,  T.
Jennewein,   C.     Simon,   H.   Weinfurter    and   A.    Zeilinger,
Phys. Rev. Lett.  {\bf 81}, 5039 (1998); P. Werbos, arXiv:0801.1234.
\bibitem{gle57} A. M.  Gleason. Measures on the closed  subspaces of a
Hilbert space.  J. Math. Mech., {\bf 6}, 885 (1957).
\bibitem{ash95} A. Peres, {\em Quantum Mechanics: Concepts and Methods},
(Kluwer, Dordrecht, 1993).
\bibitem{gis01halv}  C.    Simon,  V.   Bu{\v  z}ek   and  N.   Gisin,
Phys. Rev. lett. {\bf 87}, 170405 (2001); {\em ibid.} {\bf 90}, 208902
(2003).  H.   Halvorson, Studies in  History and Philosophy  of Modern
Physics 35, 277 (2004); quant-ph/0310101.
\bibitem{bra06} G. Brassard, H. Buhrman  and N. Linden et al.  A limit
on nonlocality in  any world in which communication  complexity is not
trivial.  Phys. Rev. Lett. {\bf 96} 250401 (2006).
\bibitem{sri06} R. Srikanth.
The quantum measurement problem and physical reality: a computation 
theoretic perspective.  AIP   Conference  Proceedings {\bf 864},
(Ed. D. Goswami) 178 (2006); quant-ph/0602114v2.
\bibitem{srirp}  In  complexity  theory,  {\bf  RP} is  the  class  of
decision  problems  for  which  there  exists a  probabilistic  TM  (a
deterministic TM with access to genuine randomness) such that: it runs
in  polynomial time  in the  input  size. If  the answer  is `no',  it
returns  `no'. If  the  answer   is  `yes',  it  returns  `yes'  with
probability at least 0.5 (else it returns `no').
{\bf  BQP} is
the  class  of  decision  problems  solvable by  a  {\em  quantum}  TM
\cite{nc00} in polynomial time, with  error probability of at most 1/3
(or,  equivalently,  any  other   fixed  fraction  smaller  than  1/2)
independently of input size.
\bibitem{nc00} M. A. Nielsen and I. Chuang, 
{\it Quantum Computation and Quantum Information}, (Cambridge 2000).
\bibitem{wit89} E. Witten. Quantum field theory and the Jones polynomial. 
Commun. Math. Phys. {\bf 121}, 351 (1989).
\bibitem{mic98} M.   Freedman. {\em P}/{\em PN} and  the quantum field
computer.  Proc. Natl. Acad. Sci. USA {\bf 95}, 98 (1998).
\bibitem{cal04} C. S.  Calude, M.   A. Stay. From Heisenberg to Goedel
via Chaitin.  Int.  J.  Th.  Phys.  44 1053 (2005); quant-ph/0402197.
\bibitem{insel} S.  Aaronson,
quant-ph/0401062; {\em ibid.},
quant-ph/0412187.
\bibitem{aar05} S. Aaronson. NP-complete Problems and Physical Reality.
ACM SIGACT News {\bf 36} 30 (2005); quant-ph/0502072. 
\bibitem{bram98}  D.  S  Abrams   and  S. Lloyd.   Nonlinear  quantum
mechanics  implies polynomial-time  solution for  NP-complete  and \#P
problems. Phys. Rev. Lett. {\bf 81}, 3992 (1998).
\bibitem{sciam} S. Aaronson. The limits of quantum computers.
Scientific American {\bf 42}, March 2008.
\bibitem{srigru}  J. Gruska. Quantum  informatics paradigms  and tools
for QIPC.  BackAction Quantum Computing: Back Action 2006, IIT Kanpur,
India,   March   2006,   Ed.    Dr.   D.   Goswami,   AIP   Conference
Proceedings. 864, pp. 1--10 (2006).
\bibitem{sriwnhe} That  is, `the universe  is not hard enough  to {\em
  not} be  simulable using  polynomial resources'.  The  expression is
  non-technically related to the  statement ``The world is not enough"
  (``{\em orbis  non sufficit}''), the family  motto of, as  well as a
  motion picture featuring, a well known Anglo-Scottish secret agent!
\bibitem{ben09} C. H. Bennett, D. Leung, G. Smith, J. A. Smolin.
Can closed timelike curves or nonlinear quantum mechanics improve quantum state 
discrimination or help solve hard problems? arXiv:0908.3023.
\bibitem{sriw}  S. Weinberg.  {\em  Dreams of  a Final  Theory}
(Vintage 1994).
\bibitem{sribpp}  More formally, {\bf  BPP} is  the class  of decision
problems solvable by a probabilistic TM in polynomial time, with error
probability of at most 1/3 (or, equivalently, any other fixed fraction
smaller than 1/2) independently of input size.
\bibitem{bag00} A. Bassi and  G. Ghirardi.  A General Argument Against
the Universal Validity of  the Superposition Principle.  Phys. Lett. A
275 (2000); quant-ph/0009020.
\bibitem{sriv1} R. Srikanth. No-signaling, intractability and
entanglement. Eprint 0809.0600.
\bibitem{ganeshpol91}  N.  Gisin, Helv.  Physica  Acta  {\bf  62}, 363  (1989);
Phys. Lett. {\bf A} 143, 1  (1990). 
J.  Polchinski.  
Phys. Rev. Lett. {\bf 66}, 397 (1991).
\bibitem{srisvet} G. Svetlichny.
Nonlinear Quantum Mechanics at the Planck Scale. Int. J. Theor. Phys. {\bf 44},
2051 (2005): quant-ph/0410230. 
\bibitem{srisvet0} G. Svetlichny. Informal Resource Letter -- 
Nonlinear quantum mechanics. quant-ph/0410036.
\bibitem{srisvet1} G. Svetlichny.  Amplification of Nonlocal Effects in 
Nonlinear Quantum Mechanics by Extreme Localization. quant-ph/0410186.
\bibitem{srisvet2} G. Svetlichny.  Nonlinear Quantum Gravity.
J.Geom.Symmetry Phys. 6 (2006) 118; quant-ph/0602012.
\bibitem{srisvet3} G. Svetlichny. Quantum Formalism with State-Collapse and 
Superluminal Communication.
Foundations of Physics, 28 131 (1998); quant-ph/9511002.
\bibitem{sen} A.  Sen-De and U.  Sen.  Testing quantum  dynamics using
signaling. Phys. Rev. A {\bf 72}, 014304 (2005).
\bibitem{pspace}  {\bf  PSPACE}  is  the class  of  decision  problems
  solvable by  a Turing  machine in polynomial  (memory) space possibly 
taking exponential time.
\bibitem{sripp}  In  complexity  theory,  {\bf  PP} is  the  class  of
decision   problems  for   which  there   exists  a   polynomial  time
probabilistic TM such  that: if the answer is  `yes', it returns `yes'
with probability greater than $1/2$, and if the answer is `no', it
returns `yes' with probability at most $1/2$.  
\bibitem{srigro97} Grover, L. K. Quantum mechanics helps in searching for
a needle in a haystack.  Phys. Rev. Lett. {\bf 79}, 325 (1997).
\bibitem{sriben97}   C.  Bennett, E.   Bernstein,  G.   Brassard  and
U.     Vazirani.    Strengths     and     weaknesses    of     quantum
computing. quant-ph/9701001.
\bibitem{sriper02} A. Peres. How the no-cloning theorem got its name.
quant-ph/0205076.
\bibitem{sri07} R.  Srikanth and S.  Banerjee. An environment-mediated
quantum deleter. Phys. Lett. A {\bf 367}, 295 (2007); quant-ph/0611161.
\bibitem{asp8} R. Ghosh and L. Mandel, Phys. Rev. Lett. {\bf 59}, 1903
(1987); P. G. Kwiat,  A. M. Steinberg and R. Y. Chiao,  Phys. Rev. {\bf A
47}, 2472 (1993); D. V. Strekalov, A. V. Sergienko, D. N. Klyshko, and
Y. H.  Shih, Phys. Rev.  Lett. {\bf 74},  3600 (1995); T.  B. Pittman,
Y. H.  Shih, D. V. Strekalov, and  A. V. Sergienko, Phys.  Rev. {\bf A
52},  R3429 (1995);  Y. -H.  Kim, R.  Yu, S.  P. Kulik,  and  Y. Shih,
M. O. Scully, Phys. Rev. Lett. {\bf 84}, 1 (2000).
\bibitem{zei00} A. Zeilinger. Experiment and foundations of quantum
physics. Rev. Mod. Phys. {\bf 71}, S288 (1999).
\bibitem{dop98}  B.  Dopfer.  Zwei  Experimente  zur  Interferenz  von
  Zwei-Photonen     Zust\"anden:    ein     Heisenbergmikroskop    und
  Pendell\"osung.  Ph.D. thesis (University of Innsbruck, 1998).
\bibitem{sri} R. Srikanth, Pramana {\bf 59}, 169 (2002); 
 {\em ibid.} quant-ph/0101023. 
\bibitem{sri888}   R.    Srikanth.    quant-ph/9904075;  {\em   ibid.}
  quant-ph/0101022.
\bibitem{sricram} J. G. Cramer, W. G. Nagourney and S. Mzali.
A test of quantum nonlocal communication. CENPA Annual Report (2007); \\
http://www.npl.washington.edu/npl/int\_rep/qm\_nl.html and
http://faculty.washington.edu/jcramer/NLS/NL\_signal.htm.
\bibitem{sridom}  R.   Jensen,  STAIF-2006  Proc.,   M.  El-Genk,
ed.                  1409                 (AIP,                 2006);
http://casimirinstitute.net/coherence/Jensen.pdf (2006).
\bibitem{srihrv} H. Nikoli\v{c}.
Is quantum field theory a genuine quantum theory? Foundational insights 
on particles and strings.  arXiv:0705.3542.
\bibitem{peres} A. Peres.  How the no-cloning theorem got its name.
arXiv:quant-ph/0205076.
\bibitem{ganeshchandru} 
C. M. Chandrashekar, Subhashish Banerjee, R. Srikanth.
Relationship Between Quantum Walk and Relativistic Quantum Mechanics.
eprint arXiv:1003.4656.
\bibitem{sridir}  We  can   then  define  Alice's  (remote)  `momentum
  observable'  as   \mbox{$\hat{P}  \equiv  f|2+5\rangle\langle2+5|  +
    f^{\prime}|1+4\rangle\langle1+4|                                  +
    f^{\prime\prime}|3+6\rangle\langle3+6|$}.     Interpreted   as   a
  quantum  field  theoretic   observable,  $\hat{P}$  is  non-complete
  because   the   projectors   to   its   eigenstates   $|2+5\rangle$,
  $|1+4\rangle$ and  $|3+6\rangle$ do not sum  to $\mathbb{I}_6 \equiv
  |1\rangle\langle1|  +   |2\rangle\langle2|  +  |3\rangle\langle3|  +
  |4\rangle\langle4|   +  |5\rangle\langle5|   +  |6\rangle\langle6|$.
  Similarly,  Alice's  non-complete  `position' observable  is  \mbox{
    $\hat{Q}        \equiv        m|1+2+3\rangle\langle1+2+3|        +
    l|4+5+6\rangle\langle4+5+6|$}.    But    note   that   $\hat{Q}$'s
  projection into  the subspace ${\cal H}_{2,5}$ is  indeed a complete
  observable.   $\hat{P}$   and  $\hat{Q}$  are  of  rank   3  and  2,
  respectively, which is smaller than 6, the dimension of the relevant
  Hilbert subspace.
\bibitem{abu01} A.F. Abouraddy, et al.
Demonstration of the complementarity of one- and two-photon interference.
Phys. Rev. A {\bf 63}, 063803 (2001).
\bibitem{bos02} S. Bose and D. Home.
Generic Entanglement Generation, Quantum Statistics, and Complementarity.
 Phys. Rev. Lett. {\bf 88}, 050401 (2002).
\bibitem{srisal08} D. Salart, A. Baas, C. Branciard, N. Gisin, and H. Zbinden.
Testing the speed of `spooky action at a distance'.
Nature 454, 861 (2008); arxiv:0808.331v1.
\bibitem{ste09} S. Skiena, {\em The Algorithm Design Manuel},
Springer (1998).
\bibitem{srizeh} H. D. Zeh. There is no "first" quantization.
Phys. Lett. A {\bf 309}, 329 (2003); quant-ph/0210098.
\bibitem{srihrv0} H. Nikoli\v{c}. There is no first quantization - 
except in the de Broglie-Bohm interpretation. quant-ph/0307179.
\bibitem{glauber} R. J. Glauber.
The Quantum Theory of Optical Coherence. Phys. Rev. 130, 2529 (1963).
\bibitem{woo} W. K. Wootters. Entanglement of Formation and Concurrence.
Quant. Info. and Comput. {\bf 1}, 27 (2001).
\end{thebibliography}
\end{document}